%% file: main.tex
\documentclass[10pt,a4paper]{article}
\usepackage{graphicx}
\usepackage[a4paper,top=20mm,bottom=23mm,left=22mm,right=22mm,headheight=15pt,footskip=11mm]{geometry}
\usepackage[T1]{fontenc}
\usepackage{lmodern}
\usepackage[scaled=0.98]{zi4}

\input{macro}

\usepackage{listings}
\usepackage{lstlinebgrd}
\usepackage{tabularx}
\usepackage{needspace}
\usepackage{microtype}
\usepackage{fancyhdr}
\usepackage{titlesec}
\usepackage{eso-pic}
\usepackage{mathtools}
\tcbuselibrary{skins,breakable,listings}

\DeclareRobustCommand{\LeanQIT}{\textnormal{\textsc{Lean-QIT}}}
\DeclareRobustCommand{\LeanQuantum}{\textnormal{\textsc{Lean-Quantum}}}
\newcommand{\snapshotDate}{July 9, 2026}

\DeclareUnicodeCharacter{00B7}{\ensuremath{\cdot}}
\definecolor{QSTNavy}{RGB}{24,55,91}
\definecolor{QSTTeal}{RGB}{20,119,128}
\definecolor{QSTInk}{RGB}{31,40,54}
\definecolor{QSTMuted}{RGB}{102,116,135}
\definecolor{QSTFrame}{RGB}{139,160,184}
\definecolor{QSTCodeBack}{RGB}{247,250,252}
\definecolor{QSTCodeGutter}{RGB}{238,245,247}
\definecolor{QSTCodeFocus}{RGB}{228,244,242}
\definecolor{QSTCodeFocusRule}{RGB}{20,119,128}
\definecolor{QSTLineNo}{RGB}{103,119,137}
\definecolor{QSTPanelBlue}{RGB}{33,94,148}
\definecolor{QSTPanelBlueDark}{RGB}{21,64,107}
\definecolor{QSTPanelBlueSoft}{RGB}{239,246,252}
\definecolor{QSTPanelBlueFrame}{RGB}{92,139,181}
\definecolor{QSTPanelCodeBack}{RGB}{247,251,255}
\definecolor{QSTPanelMeta}{RGB}{70,91,118}
\definecolor{QSTSoft}{RGB}{232,241,244}
\definecolor{QSTWarn}{RGB}{140,83,43}
\definecolor{QSTDone}{RGB}{20,119,128}
\definecolor{QITLeanDecl}{RGB}{15,77,151}
\definecolor{QITLeanNamespace}{RGB}{16,108,137}
\definecolor{QITLeanProof}{RGB}{49,118,89}
\definecolor{QITLeanType}{RGB}{36,97,153}
\definecolor{QITLeanDomain}{RGB}{38,83,128}
\definecolor{QITLeanComment}{RGB}{94,111,124}
\definecolor{QITLeanString}{RGB}{26,118,116}
\definecolor{QITLeanWarn}{RGB}{145,69,60}

\DeclareUnicodeCharacter{03A6}{\ensuremath{\Phi}}
\DeclareUnicodeCharacter{03A8}{\ensuremath{\Psi}}
\DeclareUnicodeCharacter{03B1}{\ensuremath{\alpha}}
\DeclareUnicodeCharacter{03B2}{\ensuremath{\beta}}
\DeclareUnicodeCharacter{03B3}{\ensuremath{\gamma}}
\DeclareUnicodeCharacter{03B4}{\ensuremath{\delta}}
\DeclareUnicodeCharacter{03B5}{\ensuremath{\varepsilon}}
\DeclareUnicodeCharacter{03B7}{\ensuremath{\eta}}
\DeclareUnicodeCharacter{03B8}{\ensuremath{\theta}}
\DeclareUnicodeCharacter{03B9}{\ensuremath{\iota}}
\DeclareUnicodeCharacter{03BA}{\ensuremath{\kappa}}
\DeclareUnicodeCharacter{03BB}{\ensuremath{\lambda}}
\DeclareUnicodeCharacter{03BC}{\ensuremath{\mu}}
\DeclareUnicodeCharacter{03C0}{\ensuremath{\pi}}
\DeclareUnicodeCharacter{03C1}{\ensuremath{\rho}}
\DeclareUnicodeCharacter{03C3}{\ensuremath{\sigma}}
\DeclareUnicodeCharacter{03C4}{\ensuremath{\tau}}
\DeclareUnicodeCharacter{03C6}{\ensuremath{\phi}}
\DeclareUnicodeCharacter{03C8}{\ensuremath{\psi}}
\DeclareUnicodeCharacter{03C9}{\ensuremath{\omega}}
\DeclareUnicodeCharacter{2115}{\ensuremath{\mathbb{N}}}
\DeclareUnicodeCharacter{211D}{\ensuremath{\mathbb{R}}}
\DeclareUnicodeCharacter{2192}{\ensuremath{\to}}
\DeclareUnicodeCharacter{2194}{\ensuremath{\leftrightarrow}}
\DeclareUnicodeCharacter{21A6}{\ensuremath{\mapsto}}
\DeclareUnicodeCharacter{21AA}{\ensuremath{\hookrightarrow}}
\DeclareUnicodeCharacter{2200}{\ensuremath{\forall}}
\DeclareUnicodeCharacter{2203}{\ensuremath{\exists}}
\DeclareUnicodeCharacter{2208}{\ensuremath{\in}}
\DeclareUnicodeCharacter{2209}{\ensuremath{\notin}}
\DeclareUnicodeCharacter{2211}{\ensuremath{\sum}}
\DeclareUnicodeCharacter{221E}{\ensuremath{\infty}}
\DeclareUnicodeCharacter{2227}{\ensuremath{\wedge}}
\DeclareUnicodeCharacter{2228}{\ensuremath{\vee}}
\DeclareUnicodeCharacter{2282}{\ensuremath{\subset}}
\DeclareUnicodeCharacter{2286}{\ensuremath{\subseteq}}
\DeclareUnicodeCharacter{2293}{\ensuremath{\sqcap}}
\DeclareUnicodeCharacter{2294}{\ensuremath{\sqcup}}
\DeclareUnicodeCharacter{2297}{\ensuremath{\otimes}}
\DeclareUnicodeCharacter{22A4}{\ensuremath{\top}}
\DeclareUnicodeCharacter{22A5}{\ensuremath{\bot}}
\DeclareUnicodeCharacter{2264}{\ensuremath{\le}}
\DeclareUnicodeCharacter{2265}{\ensuremath{\ge}}
\DeclareUnicodeCharacter{2260}{\ensuremath{\ne}}
\DeclareUnicodeCharacter{00B7}{\ensuremath{\cdot}}
\DeclareUnicodeCharacter{00D7}{\ensuremath{\times}}
\DeclareUnicodeCharacter{27E8}{\ensuremath{\langle}}
\DeclareUnicodeCharacter{27E9}{\ensuremath{\rangle}}

\lstdefinelanguage{LeanQIT}{
	sensitive=true,
	morekeywords=[1]{def,theorem,lemma,structure,class,instance,abbrev,inductive,noncomputable},
	morekeywords=[2]{namespace,end,variable,variables,section,open,where},
	morekeywords=[3]{by,exact,have,show,let,letI,fun,forall,exists,obtain,rcases,refine,intro,intros,rw,change,calc,from,with,inferInstance},
	morekeywords=[4]{Type,Prop,Sort,Set},
	morekeywords=[5]{axiom,sorry,admit},
	morekeywords=[6]{State,Channel,POVM,Ensemble,HypothesisTestingEffect,CMatrix,MatrixMap,SubnormalizedState,PureVector,EReal,Fintype,DecidableEq,Nonempty,Prod,BddAbove,BddBelow,IsLeast},
	morecomment=[l]{--},
	morecomment=[s]{/-}{-/},
	morestring=[b]",
}

\lstdefinestyle{QITLeanListing}{
	language=LeanQIT,
	basicstyle=\fontsize{9.8}{11.5}\selectfont\ttfamily\color{QSTInk},
	keywordstyle=[1]\color{QITLeanDecl}\bfseries,
	keywordstyle=[2]\color{QITLeanNamespace}\bfseries,
	keywordstyle=[3]\color{QITLeanProof},
	keywordstyle=[4]\color{QITLeanType}\bfseries,
	keywordstyle=[5]\color{QITLeanWarn}\bfseries,
	keywordstyle=[6]\color{QITLeanDomain}\bfseries,
	commentstyle=\itshape\color{QITLeanComment},
	stringstyle=\color{QITLeanString},
	numbers=none,
	xleftmargin=0pt,
	framexleftmargin=0pt,
	columns=fullflexible,
	keepspaces=true,
	showstringspaces=false,
	tabsize=2,
	breaklines=true,
	breakatwhitespace=false,
	breakindent=1.15em,
	breakautoindent=true,
	aboveskip=0pt,
	belowskip=0pt,
	literate=
	{_}{{\textunderscore\allowbreak}}1
	{\\Phi}{{\ensuremath{\Phi}}}1
	{\\Psi}{{\ensuremath{\Psi}}}1
	{\\alpha}{{\ensuremath{\alpha}}}1
	{\\beta}{{\ensuremath{\beta}}}1
	{\\gamma}{{\ensuremath{\gamma}}}1
	{\\delta}{{\ensuremath{\delta}}}1
	{\\epsilon}{{\ensuremath{\varepsilon}}}1
	{\\eta}{{\ensuremath{\eta}}}1
	{\\theta}{{\ensuremath{\theta}}}1
	{\\iota}{{\ensuremath{\iota}}}1
	{\\kappa}{{\ensuremath{\kappa}}}1
	{\\lambda}{{\ensuremath{\lambda}}}1
	{\\mu}{{\ensuremath{\mu}}}1
	{\\pi}{{\ensuremath{\pi}}}1
	{\\rho}{{\ensuremath{\rho}}}1
	{\\sigma}{{\ensuremath{\sigma}}}1
	{\\tau}{{\ensuremath{\tau}}}1
	{\\phi}{{\ensuremath{\phi}}}1
	{\\psi}{{\ensuremath{\psi}}}1
	{\\chi}{{\ensuremath{\chi}}}1
	{\\omega}{{\ensuremath{\omega}}}1
	{\\mathbbN}{{\ensuremath{\mathbb{N}}}}1
	{\\mathbbR}{{\ensuremath{\mathbb{R}}}}1
	{\\mathbbC}{{\ensuremath{\mathbb{C}}}}1
	{\\Rnonneg}{{\ensuremath{\mathbb{R}_{\ge 0}}}}1
	{\\to}{{\ensuremath{\to}}}1
	{\\mapsto}{{\ensuremath{\mapsto}}}1
	{\\hookrightarrow}{{\ensuremath{\hookrightarrow}}}1
	{\\forall}{{\ensuremath{\forall}}}1
	{\\exists}{{\ensuremath{\exists}}}1
	{\\infty}{{\ensuremath{\infty}}}1
	{\\in}{{\ensuremath{\in}}}1
	{\\notin}{{\ensuremath{\notin}}}1
	{\\sum}{{\ensuremath{\sum}}}1
	{\\wedge}{{\ensuremath{\wedge}}}1
	{\\vee}{{\ensuremath{\vee}}}1
	{\\otimes}{{\ensuremath{\otimes}}}1
	{\\top}{{\ensuremath{\top}}}1
	{\\bot}{{\ensuremath{\bot}}}1
	{\\le}{{\ensuremath{\le}}}1
	{\\ge}{{\ensuremath{\ge}}}1
	{\\ne}{{\ensuremath{\ne}}}1
	{\\cdot}{{\ensuremath{\cdot}}}1
	{\\langle}{{\ensuremath{\langle}}}1
	{\\rangle}{{\ensuremath{\rangle}}}1
}

\newif\ifqitLineFocused
\providecommand{\qitFocusSpec}{}

\makeatletter
\lst@Key{qitfocus}{}{\def\qitFocusSpec{#1}}
\makeatother

\newcommand{\qitHighlightSpan}[2]{%
	\ifnum\value{lstnumber}<#1\relax
	\else
		\ifnum\value{lstnumber}>#2\relax
		\else
			\qitLineFocusedtrue
		\fi
	\fi
}

\newcommand{\qitLineBackground}[3]{%
	\qitLineFocusedfalse
	\qitFocusSpec
	\ifqitLineFocused
		\color{QSTCodeFocus}%
		\rule[-#3]{#1}{\dimexpr#2+#3\relax}%
		\kern-#1%
		\color{QSTCodeFocusRule}%
		\rule[-#3]{1.35pt}{\dimexpr#2+#3\relax}%
	\fi
}

\tcbset{
	qit focus/.style={
		listing options={
			style=QITLeanListing,
			qitfocus={#1},
			linebackgroundcolor={\color{QSTPanelCodeBack}},
			linebackgroundsep=1.1mm,
			linebackgroundwidth=\dimexpr\linewidth+2.2mm\relax,
			linebackgroundheight=\dimexpr\ht\strutbox+1.1pt\relax,
			linebackgrounddepth=\dimexpr\dp\strutbox+1.1pt\relax,
			linebackgroundcmd=\qitLineBackground
		}
	}
}

\newtcblisting{qitleancode}[1][]{%
	enhanced,
	listing only,
	listing engine=listings,
	listing options={style=QITLeanListing},
	colback=QSTPanelCodeBack,
	colframe=QSTPanelBlueFrame!62,
	colbacktitle=QSTPanelBlueDark,
	coltitle=white,
	fonttitle=\small\sffamily\bfseries,
	toptitle=0.8mm,
	bottomtitle=0.8mm,
	lefttitle=1.5mm,
	titlerule=0pt,
	boxrule=0.4pt,
	arc=1.5pt,
	boxsep=0pt,
	left=1.6mm,
	right=1.6mm,
	top=1.25mm,
	bottom=1.25mm,
	before skip=0.45\baselineskip,
	after skip=0.65\baselineskip,
	#1
}

\definecolor{QudeBlue}{RGB}{34,55,199}
\definecolor{QudeViolet}{RGB}{108,49,225}
\definecolor{QudePale}{RGB}{245,247,255}

\titleformat{\section}
  {\Large\sffamily\bfseries\color{QSTPanelBlueDark}}
  {\thesection}{0.65em}{}
  [\vspace{0.15em}\color{QSTPanelBlueFrame!55}\titlerule]
\titleformat{\subsection}
  {\large\sffamily\bfseries\color{QSTNavy}}
  {\thesubsection}{0.65em}{}
\titleformat{\subsubsection}
  {\normalsize\sffamily\bfseries\color{QSTInk}}
  {\thesubsubsection}{0.65em}{}

\makeatletter
\renewcommand{\maketitle}{%
  \thispagestyle{empty}%
  \begingroup
  \setlength{\parindent}{0pt}%
  \noindent
  \begin{minipage}[t]{0.76\textwidth}
    \vspace{0pt}%
    {\raggedright\hyphenpenalty=10000\exhyphenpenalty=10000
      \sffamily\bfseries\color{QSTInk}\fontsize{23.5}{27.5}\selectfont
      \@title\par}
  \end{minipage}%
  \hfill
  \begin{minipage}[t]{0.18\textwidth}
    \vspace{0pt}\raggedleft
    \includegraphics[width=2.65cm]{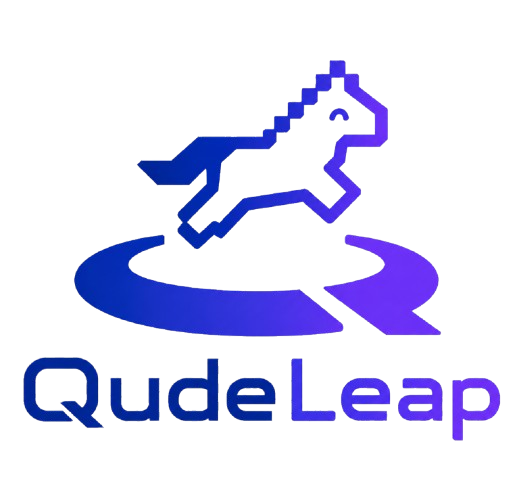}
  \end{minipage}
  \vspace{7mm}

  \noindent
  {\color{QudeBlue}\rule{0.22\textwidth}{1.8pt}}%
  {\color{QudeViolet}\rule{0.78\textwidth}{1.8pt}}\par
  \vspace{5mm}

  \begin{center}
    {\sffamily\normalsize\color{QSTInk}\@author\par}
    \vspace{2mm}
    {\small\sffamily\color{QSTMuted}\@date\par}
  \end{center}
  \vspace{3mm}
  \@thanks
  \endgroup
  \setcounter{footnote}{0}%
}
\makeatother

\title{\LeanQIT: Towards a Formal Infrastructure for \\ Quantum Information Theory}

\author[1]{Chengkai Zhu\thanks{zhuchengkai7@gmail.com}}
\author[2]{Ziao Tang\thanks{zatang004@gmail.com}}
\author[2]{Guocheng Zhen}
\author[2]{Yimeng Cao}
\author[1,2]{Yusheng Zhao}
\author[3]{Ranyiliu Chen}
\author[4,1]{Xuanqiang Zhao}
\author[1,2]{Lei Zhang\thanks{leizhang116.4@gmail.com}}
\author[2]{Xin Wang\thanks{felixxinwang@hkust-gz.edu.cn}}
\affil[1]{\small QudeLeap Research, Shanghai 200030, China}
\affil[2]{\small The Hong Kong University of Science and Technology (Guangzhou), Guangdong 511453, China}
\affil[3]{\small Quantum Science Center of Guangdong-Hong Kong-Macao Greater Bay Area, Shenzhen 518045, China}
\affil[4]{\small The University of Hong Kong, Pokfulam Road, Hong Kong}

\date{}

\begin{document}
\maketitle

\vspace{-10mm}
\begin{center}
  \small\sffamily\color{QSTMuted}%
  \raisebox{-0.5ex}{\includegraphics[height=3.2mm]{qudeleap_logo.png}}\hspace{2pt}%
  \href{https://github.com/QuAIR/Lean-QIT}{github.com/QuAIR/Lean-QIT}%
\end{center}

\begin{abstract}
Quantum information theory (QIT) characterizes the capabilities and fundamental limits of quantum information processing, underpinning quantum communication, computation, and error correction. Formalizing its coding theorems requires connecting finite-block protocols, analytic inequalities, and asymptotic limits within a unified machine-checked framework. Existing developments, however, lack a reusable operational layer that defines codes, error criteria, achievable rates, and capacities independently of their information-theoretic characterizations. In this work, we present \LeanQIT{}, a Lean 4 library for finite-dimensional QIT. It provides composable, kernel-checked interfaces for quantum states and channels, source and channel codes, finite-block performance criteria, hypothesis testing, one-shot quantities, and asymptotic rate constructions. Using this infrastructure, we formalize Schumacher's quantum source-coding theorem, the Holevo--Schumacher--Westmoreland classical capacity theorem, and the entanglement-assisted classical capacity theorem. By separating operational definitions from analytic characterizations and exposing reusable achievability, converse, and asymptotic components, \LeanQIT{} provides a machine-readable foundation for formal QIT and a compositional knowledge substrate for emerging AI-assisted formalization, automated proof search, and agentic reasoning in quantum information and computation.
\end{abstract}

\begingroup
\small
\tableofcontents
\endgroup

% ===================================================================
% \newpage
\section{Introduction}
\label{sec:intro}

Quantum information theory (QIT) studies how information is represented and processed in quantum systems, providing the foundation for quantum communication, quantum computing, and quantum error correction. Quantum Shannon theory (QST) is central to QIT. It studies the compression, transmission, and manipulation of classical and quantum information over quantum systems, and characterizes their achievable rates and fundamental reliability limits. Its principal results identify operational quantities, defined through finite-block protocols and error criteria, with entropic or information-theoretic formulae. Representative results cover source compression and communication over noisy quantum channels~\cite{Schumacher1995,Holevo1998,Schumacher1996}, quantum, private, and resource-assisted communication~\cite{Lloyd1997,Devetak2005,Bennett1996Purification,Bennett1996MixedState,Bennett1999,Bennett2002}, and channel simulation~\cite{Bennett2014QRST,Berta2011QRSTOneShot}. One-shot regimes study finite-resource settings without assuming an asymptotic repetition limit. They rely on smooth entropies, hypothesis testing, and decoupling~\cite{Renner2005,Tomamichel2015,Abeyesinghe2006,Wilde2017}. This makes QST a demanding test case for formal infrastructure in quantum information theory: quantum states, channels, information quantities, and measurements must compose with coding objects, reliability criteria, direct and converse bounds, and asymptotic limits.

The difficulty of formalizing QIT is both analytic and compositional. Coding theorems, which lie at the core of QST, must align a typed operational model with estimates across several scales. These include finite-block codes, error criteria, one-shot inequalities, tensor-power bookkeeping, direct and converse arguments, and limiting constructions. Modern proofs combine operator inequalities with finite-block coding, one-shot bounds, and symmetry or asymptotic reductions~\cite{Wilde2017,KhatriLamiWilde2025Principles}. Validity can depend on support and positivity conditions, normalization conventions, register identities, error criteria, extended-real values, and the order of quantifiers and limits. Mature informal proofs often compress these dependencies. A recently identified gap in proofs of the generalized quantum Stein's lemma~\cite{Berta2023} shows that suppressed assumptions can have substantive mathematical consequences. Formalization must therefore establish that analytic, operational, and asymptotic interfaces compose correctly.

This need becomes more pressing as AI systems increasingly enter mathematical and theoretical-physics research workflows. Their most credible near-term role in QIT is structured assistance with theorem retrieval, lemma selection, proof completion, assumption auditing, and the translation of informal arguments into formal obligations~\cite{Yang2024FormalMathReasoning,Xin2026AXLE}. Such tasks require a digital knowledge substrate with typed objects, explicit side conditions, searchable dependencies, and kernel-checkable candidate derivations. Papers with unformalized proofs alone do not provide this substrate. A reusable formal library can instead support both human researchers and proof AI agents.

Lean and its surrounding ecosystem make library-scale formalization a viable scientific methodology~\cite{deMoura2015,deMoura2021}. Mathlib has shown how advanced mathematics can be organized through reusable abstractions, stable interfaces, automation, and distributed maintenance~\cite{mathlib2020}. Domain libraries extend this model to computer science and theoretical physics~\cite{Barrett2026CSLib,PhyslibRepository,ToobySmith2024}. Lean formalizations now cover quantum programs and algorithms~\cite{Bordg2020,Peng2023}, hypothesis testing, error correction, and optimization~\cite{Meiburg2025,Ehatamm2026LeanQEC,Kol2026QuantumOptimization}. The wider Lean repository landscape spans operator theory, resource theories, and quantum foundations~\cite{Meiburg2026QuantumLeanRepos}.

We develop \LeanQIT{}~\cite{leanqit}, a Lean 4 infrastructure for finite-dimensional QIT organized into three compositional layers. The object layer provides typed states, completely positive trace-preserving (CPTP) channels, subsystem operations, classical registers, measurements, and tensor-power constructions. The analytic layer provides state geometry, purification, entropy and divergence quantities, data processing, hypothesis testing, smooth entropies, one-shot methods, typicality, and asymptotic bridges. The operational layer defines source and channel codes, error criteria, finite-block witnesses, achievable rates, capacities, strong-converse rates, and direct and converse interfaces. Its central design choice is to define operational quantities independently of the analytic formulae that later characterize them. A capacity equality is therefore a theorem relating independently specified objects, rather than a result built into the operational definition. The broader library also contains protocols, security, entanglement, and nonlocality modules, while this paper focuses on the shared substrate and its operational QST applications.

We validate this architecture through an operational QST theorem spine: Schumacher source coding, the Holevo--Schumacher--Westmoreland (HSW) classical capacity theorem, and entanglement-assisted (EA) classical capacity theorem. Together, these results cover source and channel coding, unassisted and resource-assisted communication, direct and converse reasoning, and finite-block and asymptotic methods. They also compare single-use, or single-letter, formulae with quantities regularized over many channel uses. Their proof routes require different combinations of typicality, continuity, packing, expurgation, hypothesis testing, and R\'enyi methods. We remark that, concurrent work \LeanQuantum{}~\cite{Kasaura2026LeanQuantum} develops a basis-independent operator and entropy layer and uses sandwiched R\'enyi data processing as a theorem-scale demonstration.  \LeanQIT{} targets another underdeveloped part of the formal landscape, the operational interface connecting analytic foundations to complete information-processing tasks. Together, these efforts broaden Lean support across the analytic and operational layers needed for a more comprehensive formal development of quantum information theory. Online documentation for \LeanQIT{} is available at \url{https://www.quairkit.com/Lean-QIT}.

\paragraph{Organization.}
Section~\ref{sec:architecture} describes the repository architecture and the design choices behind the formalization.  Section~\ref{sec:building-blocks} develops the reusable finite-dimensional QIT building blocks. Section~\ref{sec:coding-spine} shows how these interfaces assemble into the QST theorem spine from source coding to classical communication and entanglement assistance.  Section~\ref{sec:conclusion} discusses the resulting formalization agenda.

\paragraph{Notation and conventions.}
All Hilbert spaces in the paper are finite-dimensional.  Capital letters such
as $A,B,R$ denote quantum systems or registers, $\mathcal H_A$ denotes the
Hilbert space of system $A$, and $\mathcal L(\mathcal H_A)$ denotes its linear operators.  We write $\mathcal D(\mathcal H_A)$ for density operators on $A$. Channels such as $\mathcal N,\Phi,\Psi$ are completely positive
trace-preserving (CPTP) maps.  The symbol $\id_R$ denotes the identity channel	on register $R$.  A positive operator-valued measure (POVM) is a finite family
of positive effects that sum to the identity.  We write
$\overline{\mathbb R}=\mathbb R\cup\{-\infty,+\infty\}$ for the
extended real line.  The von Neumann entropy of a quantum state on system A is defined by $H(A)_\rho =-\tr(\rho\log\rho)$.  All logarithms in information quantities are
base two unless stated otherwise.  The quantum relative entropy is defined by
\begin{equation}
    D(\rho\Vert\sigma)=\tr\!\left[\rho(\log\rho-\log\sigma)\right],
\end{equation}
when $\operatorname{supp}(\rho)\subseteq\operatorname{supp}(\sigma)$, where
$\operatorname{supp}$ denotes the operator support, and is $+\infty$
otherwise.  The binary entropy is defined by
\begin{equation}
    h_2(x)=-x\log x-(1-x)\log(1-x),
\end{equation}
with the convention $0\log 0=0$.  We use the term application programming
interface (API) to denote a stable public collection of Lean definitions,
theorem declarations, and coercions.  Other proofs can import this collection
without depending on private proof scripts.

% ===================================================================
\section{\LeanQIT{} Architecture}
\label{sec:architecture}

\begin{figure}[ht]
\centering
\definecolor{NSNavy}{HTML}{244A73}
\definecolor{NSBlue}{HTML}{4C78A8}
\definecolor{NSInk}{HTML}{26313F}
\definecolor{NSSlate}{HTML}{5F6B7A}
\definecolor{NSLine}{HTML}{BCC8D4}
\definecolor{NSFill}{HTML}{F5F7FA}
\definecolor{NSSky}{HTML}{EAF1F8}
\definecolor{NSGreen}{HTML}{5D8B63}
\definecolor{NSGreenFill}{HTML}{EEF5EF}
\definecolor{NSOrange}{HTML}{C97A3D}
\definecolor{NSOrangeFill}{HTML}{FBF3EA}

\includegraphics[width=1\linewidth]{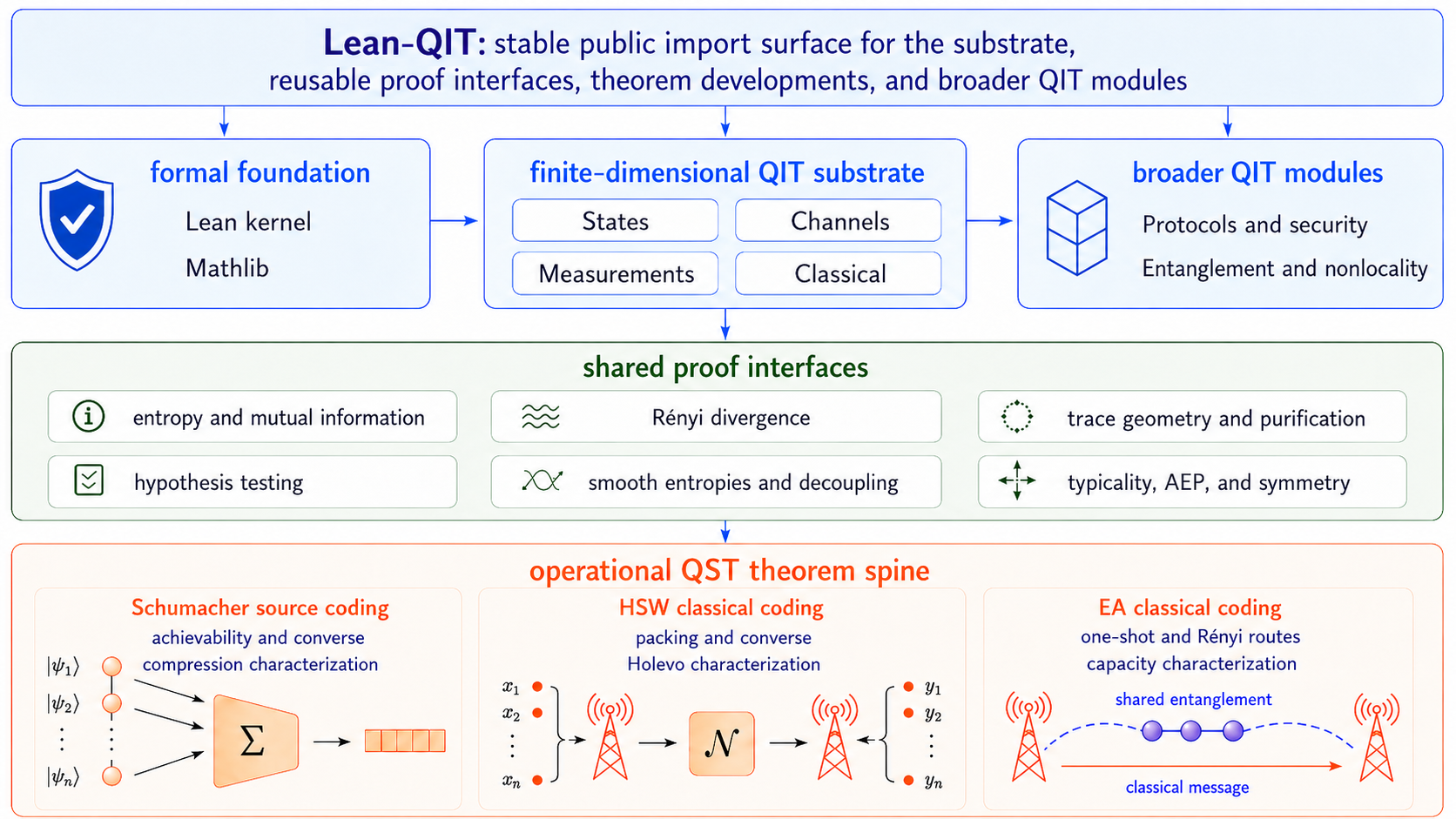}
\caption{Global architecture of \LeanQIT{}. The libaray is built on the trusted Lean/Mathlib foundation,
finite-dimensional QIT substrate, shared proof interfaces, operational QST coding surfaces, and broader QIT modules.  The QST spine consumes the shared interfaces to state and prove Schumacher source coding, HSW classical communication, and entanglement-assisted classical communication endpoints.}
\label{fig:architecture}
\end{figure}

\LeanQIT{} is organized as a proof library with stable public facades rather
than as a collection of isolated theorem examples.  The root import
\texttt{QIT.lean} re-exports finite-dimensional foundations, states,
channels, measurements, classical interfaces, information quantities,
one-shot tools, hypothesis testing, asymptotics, symmetry, coding, protocols,
and broader QIT modules.  Figure~\ref{fig:architecture} summarizes this
architecture as a global proof stack.  This section focuses
on repository organization, import surfaces, proof-status conventions, and
how code is displayed. The mathematical interfaces themselves are introduced
in Section~\ref{sec:building-blocks}.

The first architectural choice is to keep module boundaries close to the
mathematical roles used by QIT proofs.  Quantum systems are represented by
finite index types and complex matrices. State, channel, measurement,
classical-register, tensor-product, and tensor-power modules then provide the
object language used by block coding.  Information, one-shot,
hypothesis-testing, asymptotic, and symmetry modules sit above this object
layer, while the coding modules consume those APIs to assemble source-coding,
unassisted classical communication, and entanglement-assisted
classical-communication endpoints.  The broader QIT modules remain in the same public facade as additional consumers of the shared substrate. The theorem-scale path developed here is the operational QST spine.

The second architectural choice is to separate theorem status from proof-route provenance.  In the sequel, a \emph{proved public endpoint} means a declaration in the public \texttt{QIT} namespace whose proof is checked by the Lean kernel. A \emph{supporting API} is a definition or theorem used by such endpoints but not itself a final coding theorem.  A \emph{source theorem statement} is the
reader-facing mathematical claim registered from the cited QIT literature. A \emph{proof route} is the chain of identities, inequalities, reductions, and limit arguments used in Lean to establish that statement; it may follow the displayed source proof or use a mathematically equivalent alternative
derivation.  A \emph{future boundary} is an obligation that the library exposes but does not claim as completed.  This terminology separates statement alignment, proof provenance, completed endpoints, and remaining obligations. In the \snapshotDate{} snapshot, the public \LeanQIT{} library contains
$>$200 Lean files and $>$150,000 lines.
    
% ===================================================================
\section{Finite-Dimensional QIT Building Blocks}
\label{sec:building-blocks}
	
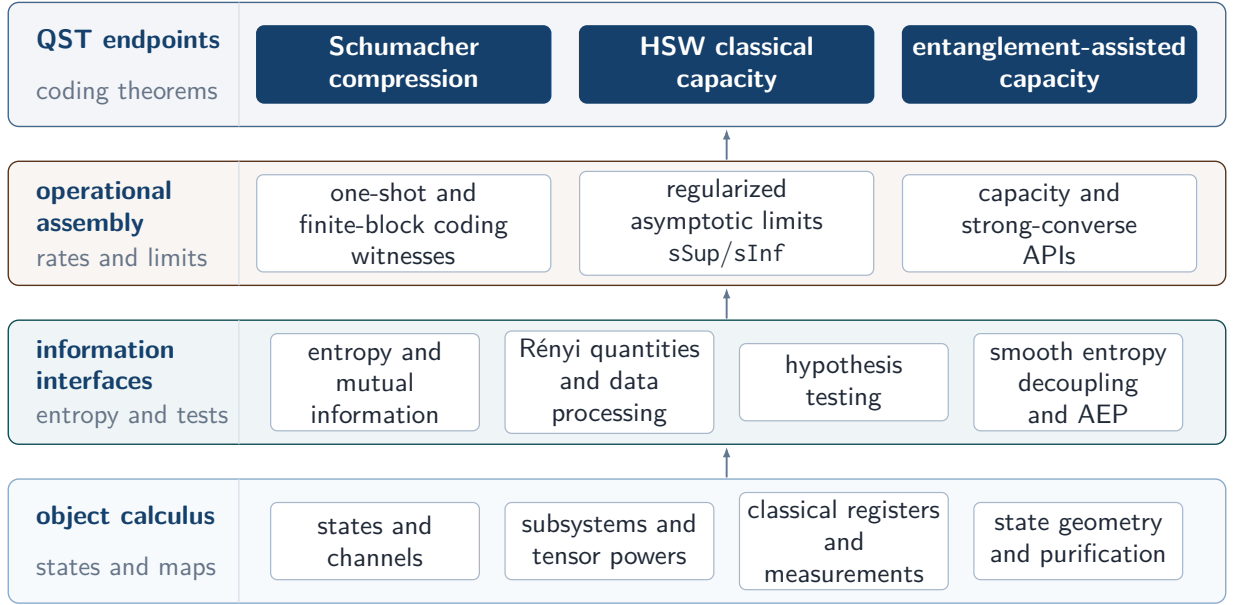
\begin{figure}[ht]
\centering
\begin{tikzpicture}[
x=1cm,
y=1cm,
every node/.style={font=\sffamily,text=QSTInk,outer sep=0pt},
layerTitle/.style={font=\normalsize\sffamily\bfseries,text=QSTPanelBlueDark,align=left,text width=2.62cm,inner sep=0pt},
layerSub/.style={font=\normalsize\sffamily,text=QSTMuted,align=left,text width=2.62cm,inner sep=0pt},
module/.style={draw=QSTFrame!75,fill=white,rounded corners=2.5pt,line width=0.46pt,text width=2.62cm,minimum height=1.02cm,align=center,inner xsep=2.0pt,inner ysep=2.8pt,font=\normalsize\sffamily,execute at begin node={\hyphenpenalty=10000\exhyphenpenalty=10000\relax}},
ratemodule/.style={draw=QSTFrame!75,fill=white,rounded corners=2.5pt,line width=0.46pt,text width=3.72cm,minimum height=1.08cm,align=center,inner xsep=2.4pt,inner ysep=2.8pt,font=\normalsize\sffamily,execute at begin node={\hyphenpenalty=10000\exhyphenpenalty=10000\relax}},
endpoint/.style={draw=QSTPanelBlueDark,fill=QSTPanelBlueDark,text=white,rounded corners=2.8pt,line width=0.50pt,text width=3.72cm,minimum height=1.02cm,align=center,inner xsep=2.4pt,inner ysep=2.8pt,font=\normalsize\sffamily\bfseries,execute at begin node={\hyphenpenalty=10000\exhyphenpenalty=10000\relax}},
layerflow/.style={-{Latex[length=1.42mm,width=1.04mm]},line width=0.72pt,draw=QSTNavy!68,line cap=round,line join=round,shorten <=0pt,shorten >=0pt}
]

\path[use as bounding box] (-8.20,-0.10) rectangle (8.20,8.02);

\draw[draw=QSTPanelBlueFrame!72,fill=QSTPanelBlueSoft!54,line width=0.52pt,rounded corners=4pt] (-8.05,-0.08) rectangle (8.05,1.56);
\draw[draw=QSTDone!64!black,fill=QSTDone!7,line width=0.52pt,rounded corners=4pt] (-8.05,2.02) rectangle (8.05,3.66);
\draw[draw=QSTWarn!62!black,fill=QSTWarn!6,line width=0.52pt,rounded corners=4pt] (-8.05,4.12) rectangle (8.05,5.76);
\draw[draw=QSTPanelBlueDark!80,fill=QSTPanelBlueDark!5,line width=0.52pt,rounded corners=4pt] (-8.05,6.22) rectangle (8.05,7.86);

\draw[draw=QSTFrame!34,line width=0.46pt] (-5.00,-0.08) -- (-5.00,1.56);
\draw[draw=QSTFrame!34,line width=0.46pt] (-5.00,2.02) -- (-5.00,3.66);
\draw[draw=QSTFrame!34,line width=0.46pt] (-5.00,4.12) -- (-5.00,5.76);
\draw[draw=QSTFrame!34,line width=0.46pt] (-5.00,6.22) -- (-5.00,7.86);

\node[layerTitle,anchor=north west] at (-7.68,1.21) {object calculus};
\node[layerSub,anchor=south west] at (-7.68,0.23) {states and maps};

\node[module] (states) at (-3.20,0.74) {states and\\channels};
\node[module] (subsys) at (-0.10,0.74) {subsystems and\\tensor powers};
\node[module] (classical) at (3.00,0.74) {classical registers\\and measurements};
\node[module] (geom) at (6.10,0.74) {state geometry\\and purification};

\node[layerTitle,anchor=north west] at (-7.68,3.41) {information\\interfaces};
\node[layerSub,anchor=south west] at (-7.68,2.23) {entropy and tests};

\node[module] (entropy) at (-3.20,2.84) {entropy and\\mutual information};
\node[module] (renyi) at (-0.10,2.84) {R\'enyi quantities\\and data processing};
\node[module] (testing) at (3.00,2.84) {hypothesis\\ testing};
\node[module] (smooth) at (6.10,2.84) {smooth entropy\\decoupling and AEP};

\node[layerTitle,anchor=north west] at (-7.68,5.51) {operational\\assembly};
\node[layerSub,anchor=south west] at (-7.68,4.37) {rates and limits};

\node[ratemodule] (witness) at (-2.82,4.94) {one-shot and\\finite-block coding\\witnesses};
\node[ratemodule] (limits) at (1.45,4.94) {regularized\\asymptotic limits\\\texttt{sSup}/\texttt{sInf}};
\node[ratemodule] (capacity) at (5.72,4.94) {capacity and\\strong-converse\\APIs};

\node[layerTitle,anchor=north west] at (-7.68,7.5) {QST endpoints};
\node[layerSub,anchor=south west] at (-7.68,6.52) {coding theorems};

\node[endpoint] (sch) at (-2.82,7.04) {Schumacher\\compression};
\node[endpoint] (hsw) at (1.45,7.04) {HSW classical\\capacity};
\node[endpoint] (ea) at (5.72,7.04) {entanglement-assisted\\capacity};

\draw[layerflow] (1.45,1.59) -- (1.45,1.99);
\draw[layerflow] (1.45,3.69) -- (1.45,4.09);
\draw[layerflow] (1.45,5.79) -- (1.45,6.19);

\end{tikzpicture}
    \caption{Layered proof stack for the reusable finite-dimensional QIT
        building blocks in \LeanQIT{}.  The object calculus supplies typed states,
        channels, subsystem operations, classical interfaces, and state geometry.
        Information-theoretic interfaces expose entropy, divergence, testing, and
        smoothing principles to operational rate constructions.  These layers are
        then assembled into the Schumacher, HSW, and entanglement-assisted coding
        endpoints developed in Section~\ref{sec:coding-spine}.}
    \label{fig:qst-proof-flow}
\end{figure}

This section explains the mathematical APIs that make the architecture of
Section~\ref{sec:architecture} usable in proofs. QST coding theorems are the most visible consumers of these interfaces, but the same APIs also support state geometry, nonlocality, and quantum information processing protocols. Figure~\ref{fig:qst-proof-flow} presents this layer as a bottom-up dependency stack. Typed states, channels, subsystem operations, classical registers, measurements, and geometric notions form the object substrate.  Entropic, R\'enyi, hypothesis-testing, smooth-entropy, decoupling, and data-processing interfaces sit above that substrate.  Rate and capacity interfaces then lift finite-block or one-shot estimates to the coding endpoints used in Section~\ref{sec:coding-spine}.

Each subsection below isolates one proof obligation that recurs across finite-dimensional QIT, e.g., forming well-typed states and channels, moving between subsystems and classical registers, measuring distances between states, applying data-processing principles, and turning finite tests or smooth entropies into asymptotic rate statements.  The coding theorems in Section~\ref{sec:coding-spine} consume these interfaces.

The listings are organized as interface cards: each title states the role of
the displayed declaration in the argument.  The main text suppresses imports,
docstrings, and surrounding namespace context; within each card, declaration
names, operational quantifiers, theorem conclusions, and displayed proof lines
follow the current library interfaces.  Pale teal bands mark the mathematical
contract consumed by the surrounding argument, while unmarked lines retain the
context needed to read that contract.  Appendix~\ref{app:endpoint-wrappers}
records the selected public endpoint wrappers in full.

\subsection{Typed states and channels}
\label{subsec:typed-states-channels}

The first interface is the standard finite-dimensional state/channel substrate.  A finite system is represented by a finite index type $a$, which should be read as a fixed orthonormal basis for a finite-dimensional Hilbert space $\mathcal H_A$. Coding theorems constantly form products, tensor powers, finite alphabets, and blocklength-indexed systems, and a basis-indexed matrix interface keeps these objects visible to Lean's kernel.  A normalized quantum state and a quantum channel are specified by the two standard conditions
\begin{equation*}
    \rho \in \mathcal D(\mathcal H_A)
    \quad\Longleftrightarrow\quad
    \rho \succeq 0,\qquad \tr \rho = 1,
\end{equation*}
where $\rho\succeq 0$ means that $\rho$ is positive semidefinite, and
\begin{equation*}
    \cN : \mathcal L(\mathcal H_A) \to \mathcal L(\mathcal H_B),\qquad
    \cN \otimes \id_R \ \text{is positive for all finite } R,\qquad
    \tr(\cN(X))=\tr(X).
\end{equation*}
Here $X\in\mathcal L(\mathcal H_A)$ is an arbitrary operator and
$R$ is an auxiliary finite-dimensional reference system.  Complete
positivity means precisely that $\cN\otimes\id_R$ is positive for every
such $R$, and trace preservation means that traces are unchanged by
$\cN$. The Lean substrate turns these mathematical side conditions into fields of the objects themselves.  The declaration \texttt{State a} bundles a complex matrix together with the proofs that it is positive semidefinite and has trace one.  The declaration \texttt{Channel a b} bundles a linear matrix map with complete positivity, trace preservation, and an explicit positivity-preserving field.  The latter is mathematically implied by complete positivity, but
storing it as a field makes downstream state-closure proofs direct and keeps
the code close to the way QIT arguments use channels.

\begin{qitleancode}[title={Core objects: states and channels},
qit focus={\qitHighlightSpan{3}{4}\qitHighlightSpan{9}{10}}]
structure State (a : Type u) [Fintype a] [DecidableEq a] where
  matrix : CMatrix a
  pos : matrix.PosSemidef
  trace_eq_one : matrix.trace = 1

structure Channel (a : Type u) (b : Type v) [Fintype a] [DecidableEq a]
    [Fintype b] [DecidableEq b] where
  map : MatrixMap a b
  completelyPositive : MatrixMap.IsCompletelyPositive map
  tracePreserving : MatrixMap.IsTracePreserving map
  mapsPositive : forall X : CMatrix a, X.PosSemidef -> (map X).PosSemidef
\end{qitleancode}
	
The typeclass assumptions are also part of the mathematical interface.
\texttt{Fintype a} supplies the finite sums used in traces, partial traces,
tensor powers, and coding alphabets. \texttt{DecidableEq a} supplies the
diagonal and matrix-entry comparisons needed by Lean.  The code records exactly which finite system it belongs to, which matrix represents it, and which kernel-checked proofs certify its normalization and positivity.

The channel action on a normalized state is a separate interface.  In
mathematical notation, if $\Phi$ is a CPTP map and $\rho$ is a state, then
$\Phi(\rho)$ is again a state.  The Lean definition makes this closure
property explicit. The output matrix is $\Phi(\rho)$, positivity is supplied
by \texttt{mapsPositive}, and trace-one normalization follows from trace
preservation and $\tr\rho=1$.
	
\begin{qitleancode}[title={Channels preserve normalized states}]
def Channel.applyState (Phi : Channel a b) (rho : State a) : State b where
  matrix := Phi.map rho.matrix
  pos := Phi.mapsPositive rho.matrix rho.pos
  trace_eq_one := by
    rw [Phi.tracePreserving rho.matrix, rho.trace_eq_one]
\end{qitleancode}
	
The positivity, trace-one normalization, complete positivity, trace preservation, and positivity preservation are stored as fields or closure lemmas.  Later information-theoretic statements can therefore mention $\Phi(\rho)$ as a state without repeatedly reconstructing the normalization proof, which is especially useful for agent-assisted proof
search.
	
\subsection{Subsystem, classical, and measurement calculus}
\label{subsec:subsystem-classical-measurement}

The next substrate operations are products, marginals, product channels, and
tensor powers, i.e., $\rho_A\otimes\sigma_B$, $\rho_A=\tr_B\rho_{AB}$,
$\id_R\otimes\cN$, and $\cN^{\otimes n}$. These are the bookkeeping operations that appear in almost every QST proof.  \LeanQIT{} exposes them as
typed operations, so that later entropy and coding lemmas can talk about the
same objects rather than reimplementing subsystem bookkeeping.  The same
layer also supplies classical registers, ensembles, classical--quantum (cq)
states, and POVM measurement outputs, which are indispensable for HSW coding, hypothesis testing, and classical post-processing.

\Needspace{20\baselineskip}
\begin{qitleancode}[title={Product states and marginals}]
def State.prod (rho : State a) (sigma : State b) : State (Prod a b) where
  matrix := Matrix.kronecker rho.matrix sigma.matrix
  pos := rho.pos.kronecker sigma.pos
  trace_eq_one := by
    change (Matrix.kroneckerMap (fun x y => x * y) rho.matrix sigma.matrix).trace = 1
    rw [Matrix.trace_kronecker, rho.trace_eq_one, sigma.trace_eq_one]
    norm_num

def State.marginalA (rho : State (Prod a b)) : State a where
  matrix := partialTraceB (a := a) (b := b) rho.matrix
  pos := partialTraceB_posSemidef rho.pos
  trace_eq_one := by
    rw [partialTraceB_trace, rho.trace_eq_one]
\end{qitleancode}
	
The first two declarations are state-level subsystem operations.  The product
state constructs $\rho_A\otimes\sigma_B$ and carries the positivity and
trace-one proofs for the tensor product.  The marginal operation constructs
$\tr_B\rho_{AB}$ and carries the proof that partial trace preserves
positivity and normalization.  Product channels are then connected to these
state operations by a separate compatibility theorem.

\Needspace{12\baselineskip}
\begin{qitleancode}[title={Product-channel action}]
theorem Channel.applyState_prod (Phi : Channel a b) (Psi : Channel c d)
    (rho : State a) (sigma : State c) :
    (Phi.prod Psi).applyState (rho.prod sigma) =
      (Phi.applyState rho).prod (Psi.applyState sigma) := by
  apply State.ext
  change (Phi.prod Psi).map (Matrix.kronecker rho.matrix sigma.matrix) =
    Matrix.kronecker (Phi.map rho.matrix) (Psi.map sigma.matrix)
  exact prod_map_kronecker Phi Psi rho.matrix sigma.matrix
\end{qitleancode}
	
The displayed theorem is the formal version of the identity
\begin{equation}
    (\Phi\otimes\Psi)(\rho\otimes\sigma)
    =
    \Phi(\rho)\otimes\Psi(\sigma).
\end{equation}
Its role is larger than this elementary formula suggests.  It is the typed
bridge behind output states such as $(\id_R\otimes\cN)(\psi_{RA})$, behind
tensor-power channels $\cN^{\otimes n}$, and behind additivity arguments in
source coding, HSW, and entanglement-assisted capacity proofs.

Classical labels and measurements are represented in the same finite
substrate. This design keeps classical--quantum states, POVMs, and binary tests in the same typed world as states and channels.
	
\begin{qitleancode}[title={Classical and measurement objects}]
structure POVM (x : Type u) (a : Type v) [Fintype x] [Fintype a] [DecidableEq a] where
  effects : x \to CMatrix a
  pos : \forall y, (effects y).PosSemidef
  sum_eq_one : \sum y, effects y = 1

structure Ensemble (\iota : Type u) (a : Type v)
    [Fintype \iota] [Fintype a] [DecidableEq a] where
  probs : \iota \to \Rnonneg
  weights_sum : (\sum i, probs i) = 1
  states : \iota \to State a
\end{qitleancode}
	
The measurement bridge is a separate classicalization operation.  Given a POVM $M$ and an input state $\rho$, the library first obtains each outcome probability through \texttt{M.prob rho outcome}; it then packages the resulting probability vector as a diagonal classical state.  This makes the post-measurement classical outcome compatible with the same \texttt{State} API used by entropy, hypothesis testing, and coding definitions.
	
\begin{qitleancode}[title={Measurement-to-classical bridge}]
namespace Classical

def measuredState (M : POVM y a) (rho : State a) : State y :=
  diagonalState (fun outcome => M.prob rho outcome) (M.sum_prob rho)

end Classical
\end{qitleancode}
	
	This bridge lets an HSW code contain a finite message type and a
	decoding POVM, while its performance is still a statement about states.  It
	also supports the hypothesis-testing layer below: the same effect calculus can
	express a POVM outcome, a decoder acceptance event, or a binary test in a
	converse proof.
	
	\subsection{State geometry and purification interfaces}
	\label{subsec:state-geometry}
	
	Operational QIT statements compare ideal and implemented states.  The
	geometry layer therefore exposes trace distance, fidelity, purified distance,
	and purification-based constructions as state-level interfaces.  For states
	$\rho,\sigma$, the normalized trace distance is
	$\frac12\|\rho-\sigma\|_1$, where $\|\cdot\|_1$ is the trace norm.  A
	basic theorem is contraction of trace distance under CPTP maps~\cite{Wilde2017}:
	\begin{equation}
		\frac12\|\Phi(\rho)-\Phi(\sigma)\|_1
		\le
		\frac12\|\rho-\sigma\|_1 .
	\end{equation}
	In \LeanQIT{}, the public theorem
	\texttt{Channel.normalizedTraceDistance\_applyState\_le} states this
	contraction directly for the state-level channel action.  Its proof extracts a
	Kraus representation, pulls the positive spectral projector of the output
	difference back to an input effect, and applies trace-norm duality.  The theorem
	is used whenever a protocol error is pushed through an encoder, decoder, local
	operation, or tensor-product channel without changing the mathematical object
	being compared.
	
	Smooth one-shot information theory uses purified distance rather than trace
	distance as its default smoothing metric~\cite{Tomamichel2015,Wilde2017}. To avoid the two common fidelity conventions, we write $f(\rho,\sigma) \coloneqq \left\|\sqrt{\rho}\sqrt{\sigma}\right\|_1$ for the root fidelity and $F(\rho,\sigma)\coloneqq f(\rho,\sigma)^2$ for the squared fidelity. Lean-QIT follows this distinction through \texttt{State.fidelity} and \texttt{State.squaredFidelity}. For normalized states, the purified distance is defined by
	\begin{equation}
		P(\rho,\sigma) \coloneqq\sqrt{1-F(\rho,\sigma)}
	\end{equation}
	and exposes the closed $\ve$-ball as a predicate on states. Later smooth entropy definitions quantify over this predicate rather than repeating the metric inequality in every theorem statement.
	
\begin{qitleancode}[title={Purified distance and smoothing}]
namespace State

def purifiedDistance (\rho \sigma : State a) : \mathbbR :=
  Real.sqrt (1 - \rho.squaredFidelity \sigma)

def purifiedBall (\rho : State a) (\epsilon : \mathbbR) (\sigma : State a) : Prop :=
  \rho.purifiedDistance \sigma \le \epsilon

end State
\end{qitleancode}
	
	This geometric layer also contains purification-specific interfaces, including
	canonical purifications and Uhlmann-style fidelity statements~\cite{Tomamichel2015,Wilde2017}.  Those results
	are not displayed here, but they explain why the same state-geometry API can
	serve Schumacher fidelity estimates, one-shot smoothing, and decoupling
	arguments.
	
	\subsection{Information quantities and data processing}
	\label{subsec:information-dpi}
	
	The information layer turns state-level constructions into entropy and divergence quantities that can be reused by converse, achievability, and strong-converse arguments.  For a bipartite state $\rho_{AB}$, with marginals $\rho_A=\tr_B\rho_{AB}$ and $\rho_B=\tr_A\rho_{AB}$, the
	quantum mutual information~\cite{Wilde2017} is defined by
	\begin{equation}
		I(A;B)_\rho \coloneqq S(\rho_A)+S(\rho_B)-S(\rho_{AB}).
	\end{equation}
	In Lean, this is a state-level definition over a product system.  The
	marginals are obtained from the subsystem API above, and
	\texttt{State.vonNeumann} is the von Neumann entropy used throughout the
	coding layer.
	
\begin{qitleancode}[title={Mutual information}]
def mutualInformation [DecidableEq \iota] (\rho : State (Prod \iota a)) : \mathbbR :=
  State.vonNeumann \rho.marginalA + State.vonNeumann \rho.marginalB
    - State.vonNeumann \rho
\end{qitleancode}
	
	The corresponding structural theorem is the data-processing inequality~\cite{Wilde2017}.  For
	local quantum channels $\Phi:A\to A'$ and $\Psi:B\to B'$, \LeanQIT{}
	states the monotonicity in the form
	\begin{equation}
		I(A;B)_\rho
		\ge
		I(A';B')_{(\Phi\otimes\Psi)(\rho)} .
	\end{equation}
	This lemma is one of the workhorses behind converse arguments, which lets a decoder or local post-processing step only decrease correlations.
	
\begin{qitleancode}[title={Data processing under local channels},
qit focus={\qitHighlightSpan{6}{7}}]
theorem mutualInformation_dataProcessing_local_channels_ge
    {a : Type u} {b : Type v} {c : Type w} {d : Type x}
    [Fintype a] [DecidableEq a] [Fintype b] [DecidableEq b]
    [Fintype c] [DecidableEq c] [Fintype d] [DecidableEq d]
    (\rho : State (Prod a b)) (\Phi : Channel a c) (\Psi : Channel b d) :
    mutualInformation \rho \ge
      mutualInformation ((\Phi.prod \Psi).applyState \rho) := by
  exact
    mutualInformation_dataProcessing_local_channels_ge_of_traceLog_bridge \rho \Phi \Psi
      (State.relativeEntropyPSDReferenceTraceLogE_prod_marginals_eq_mutualInformation \rho)
      (State.relativeEntropyPSDReferenceTraceLogE_prod_marginals_eq_mutualInformation
        ((\Phi.prod \Psi).applyState \rho))
\end{qitleancode}
	
The displayed implementation follows a source-supported relative-entropy derivation.  It identifies mutual information with $D(\rho_{AB}\Vert\rho_A\otimes\rho_B)$, applies the independently established data-processing inequality for quantum relative entropy to
$\Phi\otimes\Psi$, and verifies that the product reference is mapped to the
product of the output marginals.  The proof therefore does not depend on
strong subadditivity and introduces no circular dependency. The same pattern extends to R\'enyi quantities.  The sandwiched R\'enyi divergence is central in strong converses and one-shot bounds.  For normalized states $\rho,\sigma\succ0$ and $\alpha\in(0,1)\cup(1,\infty)$, the sandwiched R\'enyi divergence is defined by~\cite{M_ller_Lennert_2013,Wilde2014}
	\begin{equation}
			\widetilde D_\alpha(\rho\Vert\sigma)
			\coloneqq\frac{1}{\alpha-1}
			\log\tr\!\left[
			\left(\sigma^{(1-\alpha)/(2\alpha)}
			\rho\sigma^{(1-\alpha)/(2\alpha)}\right)^\alpha
			\right].
	\end{equation}
	
\begin{qitleancode}[title={Sandwiched R\'enyi divergence}]
namespace State

def sandwichedRenyi (\rho \sigma : State a)
    (_h\rho : \rho.matrix.PosDef) (_h\sigma : \sigma.matrix.PosDef)
    (\alpha : \mathbbR) (_h\alpha_pos : 0 < \alpha)
    (_h\alpha_ne_one : \alpha \ne 1) : \mathbbR :=
  let r := 1 / (\alpha - 1)
  let s := (1 - \alpha) / (2 * \alpha)
  let C := CFC.rpow \sigma.matrix s
  let M := CFC.rpow (C * \rho.matrix * C) \alpha
  r * log2 (M.trace.re)

end State
\end{qitleancode}
	
	Most QST converse proofs, however, use a second argument that is only known to
	be positive semidefinite (PSD), not necessarily a normalized full-rank state.
	\LeanQIT{} therefore exposes a source-facing extended-real PSD-reference
	definition.  The declaration below assembles the low-$\alpha$ branch and the
	support-aware high-$\alpha$ branch into a single endpoint.
	
\begin{qitleancode}[title={PSD-reference R\'enyi divergence}]
namespace State

noncomputable def sandwichedRenyiPSDReferenceE
    (\rho : State a) (\sigma : CMatrix a) (h\sigma : \sigma.PosSemidef)
    (\alpha : \mathbbR) : EReal :=
  if \alpha < 1 then
    sandwichedRenyiPSDReferenceLowAlphaE \rho \sigma h\sigma \alpha
  else
    sandwichedRenyiPSDReferenceHighAlphaE \rho \sigma h\sigma \alpha

end State
\end{qitleancode}
	
	The PSD-reference API is the one used by the data-processing inequality in the range $\alpha\in[1/2,1)\cup(1,\infty)$:
	\begin{equation}
		\widetilde D_\alpha(\rho\Vert\sigma)
		\ge
		\widetilde D_\alpha(\Phi(\rho)\Vert\Phi(\sigma)).
	\end{equation}
	The Lean statement keeps the positive semidefinite reference operator
	explicit, because converse and strong-converse proofs often apply the
	inequality before the reference has been normalized to a state.
	
\begin{qitleancode}[title={PSD-reference data processing},
qit focus={\qitHighlightSpan{8}{11}}]
namespace State

theorem sandwichedRenyiPSDReferenceE_dataProcessing_channel_ge_of_half_le_lt_one_or_one_lt
    {b : Type v} [Fintype b] [DecidableEq b]
    (\rho : State a) {\sigma : CMatrix a} (h\sigma : \sigma.PosSemidef)
    (\Phi : Channel a b) (\alpha : \mathbbR)
    (h\alpha_range : (1 / 2 \le \alpha \wedge \alpha < 1) \vee 1 < \alpha) :
    sandwichedRenyiPSDReferenceE \rho \sigma h\sigma \alpha \ge
      sandwichedRenyiPSDReferenceE
        (\Phi.applyState \rho) (\Phi.map \sigma)
        (\Phi.mapsPositive \sigma h\sigma) \alpha := by
  exact
    sandwichedRenyiPSDReferenceE_dataProcessing_channel_of_half_le_lt_one_or_one_lt
      \rho h\sigma \Phi \alpha h\alpha_range

end State
\end{qitleancode}
	
The proof route is worth recording because it explains why this theorem is an
API endpoint rather than an isolated inequality.  For a state
$\rho$ and a PSD reference operator $\sigma\succeq0$, the branch
$1/2<\alpha<1$ uses the sandwiched trace functional defined by
\begin{equation}
    Q_{\alpha}(\rho,\sigma)
    \coloneqq
    \tr\!\left[
    \left(
    \sigma^{(1-\alpha)/(2\alpha)}
    \rho
    \sigma^{(1-\alpha)/(2\alpha)}
    \right)^{\alpha}
    \right],
\end{equation}
so that the finite logarithmic branch is defined by
\begin{equation}
    \widetilde D_{\alpha}(\rho\Vert\sigma) \coloneqq
    \frac{1}{\alpha-1}\log Q_{\alpha}(\rho,\sigma).
\end{equation}
\LeanQIT{} also records the singular case $Q_{\alpha}(\rho,\sigma)=0$ in
$\overline{\mathbb R}$, where the low-$\alpha$ expression is $+\infty$. For $1/2<\alpha<1$, the proof derives monotonicity of $Q_\alpha$ from the
Frank--Lieb joint-concavity route, implemented through a variational formulation, finite unitary twirling, and a Stinespring reduction~\cite{FrankLieb2013Renyi}.  The negative factor $1/(\alpha-1)$ reverses the $Q_\alpha$ inequality when it is converted to a divergence inequality, while the case $Q_\alpha=0$ is represented by $+\infty$ in $\overline{\mathbb R}$.  At $\alpha=1/2$, Lean normalizes a nonzero PSD
reference, derives root-fidelity monotonicity from squared-fidelity
monotonicity, and separately closes the zero-reference branch.  For
$\alpha>1$, the proof uses Beigi's weighted-Schatten interpolation argument
and then treats singular references by an explicit support split: unsupported
inputs give $+\infty$, whereas supported state-reference pairs are compressed
to their positive spectral supports~\cite{Beigi2013Sandwiched}.

Thus, \LeanQIT{} establishes the data-processing inequality of sandwiched R\'enyi divergence through a branchwise alternative route rather than reproducing the source optimizer calculation and double-regularization argument.  Independently,
\LeanQuantum{} provides a Lean formalization of sandwiched R\'enyi data processing inequality within a basis-independent operator framework, thereby establishing an important analytic foundation for the formal development of generalized quantum Stein's lemma~\cite{Kasaura2026LeanQuantum}.
\LeanQIT{} develops a distinct downstream role for this inequality by exposing
its PSD-reference formulation as a reusable interface for operational converse
and strong-converse arguments in Section~\ref{sec:coding-spine}. Together,
the two developments demonstrate how a central information-theoretic
inequality can support both general operator-theoretic foundations and
operational quantum Shannon theory.
    % This is the PSD $Q_{\alpha}$-functional route associated with Gour and
	% Frank--Lieb~\cite{FrankLieb2013Renyi}. The endpoint $\alpha=1/2$ is supplied
	% by normalized reference fidelity monotonicity.  The high-$\alpha$ branch uses a
	% support-aware Beigi weighted-Schatten route~\cite{Beigi2013Sandwiched}.  The
	% same DPI is the central theorem-scale demonstration of the concurrent and
	% independent \LeanQuantum{} development, where it validates a basis-independent
	% operator-theoretic and entropic infrastructure~\cite{Kasaura2026LeanQuantum}.
	% In \LeanQIT{}, the public DPI has a complementary downstream role: it is
	% consumed by operational converse and strong-converse interfaces in
	% Section~\ref{sec:coding-spine}.
	
	% These lemmas are not tied to a single theorem in Section~\ref{sec:coding-spine}.  They are shared QIT interfaces. Schumacher converses use trace-distance and mutual-information continuity,	HSW converses use Holevo and mutual-information bounds, and entanglement-assisted strong converses use R\'enyi data processing and optimized mutual-information quantities.
	
\subsection{Testing interfaces for finite-block converses}
\label{subsec:converse-testing}

The next layer provides the proof grammar used to turn reliable protocols into
information inequalities.  Converse proofs repeatedly combine data processing,
continuity estimates, cq/Holevo identities, and dimension bounds. Finite-block
or strong-converse arguments additionally use binary tests and R\'enyi-type
exponents.  \LeanQIT{} separates these ingredients so that a coding theorem can	cite the relevant interface rather than restating the whole analytic
reduction.

The hypothesis-testing layer records one such source-level information quantity before exposing its Lean optimization interface.  As used in
finite-resource quantum information theory~\cite{Tomamichel2015,Wilde2017}, for states $\rho,\sigma$ on the same system and error tolerance $0<\ve<1$, the hypothesis-testing relative entropy is defined by
\begin{equation}
        D_{\mathrm H}^{\ve}(\rho\Vert\sigma)
        \coloneqq-\log
        \inf_{\substack{0\le \Lambda \le I\\
                \tr(\Lambda\rho)\ge 1-\ve}}
        \tr(\Lambda\sigma),
\end{equation}
where $\Lambda$ is a binary testing effect, $I$ is the identity operator, and the constraint $\tr(\Lambda\rho)\ge1-\ve$ bounds the type-I error by $\ve$. For a bipartite state $\rho_{AB}$, the optimized hypothesis-testing mutual information is defined by
\begin{equation}
    I_{\mathrm H}^{\ve}(A;B)_\rho
    \coloneqq
    \inf_{\sigma_B \in \mathcal D(\mathcal H_B)}
    D_{\mathrm H}^{\ve}
    \bigl(\rho_{AB}\Vert \rho_A\otimes\sigma_B\bigr),
\end{equation}
where the infimum is over density operators $\sigma_B$ on system $B$.
The channel quantity is defined by optimizing over input-reference pure states
\begin{equation}
        I_{\mathrm H}^{\ve}(\cN)
        \coloneqq
        \sup_{\psi_{RA}}
        I_{\mathrm H}^{\ve}(R;B)_
        {(\id_R\otimes \cN)(\psi)},
\end{equation}
where $\psi_{RA}$ ranges over pure states on a reference $R$ and channel
input $A$, and $B$ denotes the output system of $\cN$.
The binary testing interface is the kernel-level witness for the constraint
$0\le\Lambda\le I$ and $\tr(\Lambda\rho)\ge 1-\ve$.
	
\begin{qitleancode}[title={Feasible hypothesis-testing effects},
qit focus={\qitHighlightSpan{3}{5}}]
structure HypothesisTestingEffect (rho : State a) (epsilon : \mathbbR) where
  effect : CMatrix a
  pos : effect.PosSemidef
  le_one : effect \le 1
  accept_ge : 1 - epsilon \le effectAcceptProbability rho effect
\end{qitleancode}
	
	At the state level, Lean represents the outer optimization by a candidate set
	and an infimum.  The candidate set ranges over auxiliary states
	$\sigma_B$, and \texttt{sInf} records the optimized value
	$I_{\mathrm H}^{\ve}(A;B)_\rho$ as an order-theoretic real quantity.
	
\begin{qitleancode}[title={State-level hypothesis-testing information}]
namespace State

def hypothesisTestingMutualInformationCandidateSet
    (rhoAB : State (Prod a b)) (epsilon : \mathbbR) : Set \mathbbR :=
  {value | \exists sigmaB : State b,
    value = rhoAB.hypothesisTestingRelativeEntropy
      (rhoAB.marginalA.prod sigmaB) epsilon}

def hypothesisTestingMutualInformationECandidateSet
    (rhoAB : State (Prod a b)) (epsilon : \mathbbR) : Set EReal :=
  {value | \exists sigmaB : State b,
    value = rhoAB.hypothesisTestingRelativeEntropyE
      (rhoAB.marginalA.prod sigmaB) epsilon}

def hypothesisTestingMutualInformation
    (rhoAB : State (Prod a b)) (epsilon : \mathbbR) : \mathbbR :=
  sInf (hypothesisTestingMutualInformationCandidateSet (a := a) (b := b) rhoAB epsilon)

def hypothesisTestingMutualInformationE
    (rhoAB : State (Prod a b)) (epsilon : \mathbbR) : EReal :=
  sInf (hypothesisTestingMutualInformationECandidateSet (a := a) (b := b) rhoAB epsilon)

end State
\end{qitleancode}
	
The channel-level interface is displayed separately because it adds a
different kind of optimization.  First, the channel acts on an
input-reference pure state through $(\id_R\otimes\cN)(\psi_{RA})$.  Then the
channel hypothesis-testing mutual information is the supremum over such
input-reference states.  The real-valued quantity is convenient for ordinary
finite-block estimates, while the extended-real variant is used by one-shot
and R\'enyi strong-converse routes.
	
\begin{qitleancode}[title={Channel-level hypothesis-testing information}]
namespace Channel

def hypothesisTestingOutputState {r : Type w} [Fintype r] [DecidableEq r]
    (\psi : PureVector (Prod r a)) : State (Prod r b) :=
  ((Channel.idChannel r).prod N).applyState \psi.state

def hypothesisTestingMutualInformationValueSet (epsilon : \mathbbR) : Set \mathbbR :=
  Set.range fun \psi : PureVector (Prod a a) =>
    N.inputHypothesisTestingMutualInformation \psi epsilon

def hypothesisTestingMutualInformation (epsilon : \mathbbR) : \mathbbR :=
  sSup (N.hypothesisTestingMutualInformationValueSet epsilon)

def hypothesisTestingMutualInformationEValueSet (epsilon : \mathbbR) : Set EReal :=
  Set.range fun \psi : PureVector (Prod a a) =>
    N.inputHypothesisTestingMutualInformationE \psi epsilon

def hypothesisTestingMutualInformationE (epsilon : \mathbbR) : EReal :=
  sSup (N.hypothesisTestingMutualInformationEValueSet epsilon)

end Channel
\end{qitleancode}
	
Together these two code blocks show a QIT-specific interface rather than a
coding-specific endpoint.  A paper-facing operational quantity is decomposed
into a state-level infimum, a channel output-state construction, and a
channel-level supremum.  Coding theorems can then cite the optimized quantity
without restating the binary-testing optimization each time.

\subsection{One-shot and asymptotic bridges}
\label{subsec:smooth-asymptotic}
	
For the one-shot layer, the displayed source definition is the smoothed
conditional min-entropy.  In a standard finite-dimensional convention for
smooth entropies~\cite{Renner2005,Tomamichel2015}, for a bipartite state
$\rho_{AB}$, the conditional min-entropy is defined by
\begin{equation}
        H_{\min}(A|B)_\rho
        \coloneqq
        -\log
        \inf_{\sigma_B\in\mathcal D(\mathcal H_B)}
        \inf\{\lambda:\rho_{AB}\le \lambda I_A\otimes\sigma_B\}.
\end{equation}
Let $\mathcal S_{\le}(\mathcal H)\coloneqq \{\tau\succeq 0:\operatorname{Tr}\tau\le 1\}$. For \(\tau,\omega\in\mathcal S_{\le}(\mathcal H)\), define the generalized root fidelity by
\begin{equation}
f_*(\tau,\omega)\coloneqq \left\|\sqrt{\tau}\sqrt{\omega}\right\|_1 + \sqrt{(1-\operatorname{Tr}\tau)(1-\operatorname{Tr}\omega)},
\end{equation}
and let \(F_*(\tau,\omega):=f_*(\tau,\omega)^2\). The purified distance is $P(\tau,\omega)\coloneqq\sqrt{1-F_*(\tau,\omega)}$. Accordingly, $\mathcal B^\varepsilon(\rho) \coloneqq \{\widetilde\rho\in\mathcal S_{\le}(\mathcal H): P(\widetilde\rho,\rho)\le\varepsilon\},~0\le\varepsilon<1$.
The smoothed conditional min-entropy is defined by
\begin{equation}
    H_{\min}^{\ve}(A|B)_\rho \coloneqq
    \sup_{\tilde\rho\in\mathcal B^{\ve}(\rho)}
    H_{\min}(A|B)_{\tilde\rho}.
\end{equation}
The asymptotic interface targets the standard fully quantum AEP for smooth conditional entropies.

\begin{theorem}[Fully quantum asymptotic equipartition
property~\cite{TomamichelColbeckRenner2009AEP}]
\label{thm:fully-quantum-aep}
Let $\rho_{AB}$ be a bipartite state on finite-dimensional systems
$A$ and $B$. Then
\begin{align}
    \lim_{\ve\to0^+}\lim_{n\to\infty}
    \frac{1}{n}
    H_{\min}^{\ve}(A^n|B^n)_{\rho_{AB}^{\otimes n}}
    &= H(A|B)_\rho, \\
    \lim_{\ve\to0^+}\lim_{n\to\infty}
    \frac{1}{n}
    H_{\max}^{\ve}(A^n|B^n)_{\rho_{AB}^{\otimes n}}
    &= H(A|B)_\rho .
\end{align}
\end{theorem}
	
The code block below is not the AEP itself.  It packages a \LeanQIT{}-specific candidate-lifting lemma for the optimization defining smooth conditional min-entropy, whose standard finite-dimensional formulation is developed in Refs.~\cite{Renner2005,Tomamichel2015}.  Because the entropy is represented as a supremum over feasible candidates, lifting every postprocessed candidate to a source-side candidate with no smaller value yields the corresponding entropy inequality.  The convenience theorem keeps this candidate transport explicit while discharging nonemptiness and boundedness through the smooth-entropy API.
	
\begin{qitleancode}[title={Smooth-entropy comparison interface},
qit focus={\qitHighlightSpan{12}{13}}]
theorem smoothConditionalMinEntropy_le_of_candidate_lift_of_lt_sqrt_trace
    {source : Type w} [Fintype source] [DecidableEq source]
    [Nonempty source] [Nonempty b]
    (\rhoPost : SubnormalizedState (Prod a b))
    (\rhoSource : SubnormalizedState (Prod source b)) {\epsilon : \mathbbR}
    (h\epsilon0 : 0 \le \epsilon)
    (h\epsilonSource : \epsilon < Real.sqrt \rhoSource.matrix.trace.re)
    (hlift : \forall h,
      SmoothConditionalMinEntropyCandidate (a := a) \rhoPost \epsilon h \to
        \exists h',
          SmoothConditionalMinEntropyCandidate (a := source) \rhoSource \epsilon h' \wedge h \le h') :
    \rhoPost.smoothConditionalMinEntropy \epsilon \le
      \rhoSource.smoothConditionalMinEntropy \epsilon :=
  smoothConditionalMinEntropy_le_of_candidate_lift \rhoPost \rhoSource
    (SmoothConditionalMinEntropyCandidate_set_nonempty_of_nonneg (a := a) \rhoPost h\epsilon0)
    (SmoothConditionalMinEntropyCandidate_bddAbove_of_lt_sqrt_trace
      (a := source) \rhoSource h\epsilonSource)
    hlift
\end{qitleancode}
	
The displayed candidate-lifting theorem is a reusable neighboring interface. The public AEP endpoint is assembled from the finite-$N$ AEP lower bound~\cite{TomamichelColbeckRenner2009AEP}, the ordering inequality
$H_{\min}\leq H$~\cite{Tomamichel2015}, the Alicki--Fannes--Winter continuity
bound~\cite{AlickiFannes2004,Winter2016EntropyContinuity}, smooth
min/max-entropy duality~\cite{TomamichelColbeckRenner2010Duality}, and
conditional von Neumann entropy duality~\cite{Wilde2017}. These named interfaces yield the two-stage fully quantum AEP~\cite{TomamichelColbeckRenner2009AEP}.  Its public wrapper performs only finite-type nonemptiness plumbing and forwards the established analytic endpoints, as illustrated in Appendix~\ref{app:endpoint-wrappers}.

% ===================================================================
\section{Coding Spine of Quantum Shannon Theory}
\label{sec:coding-spine}

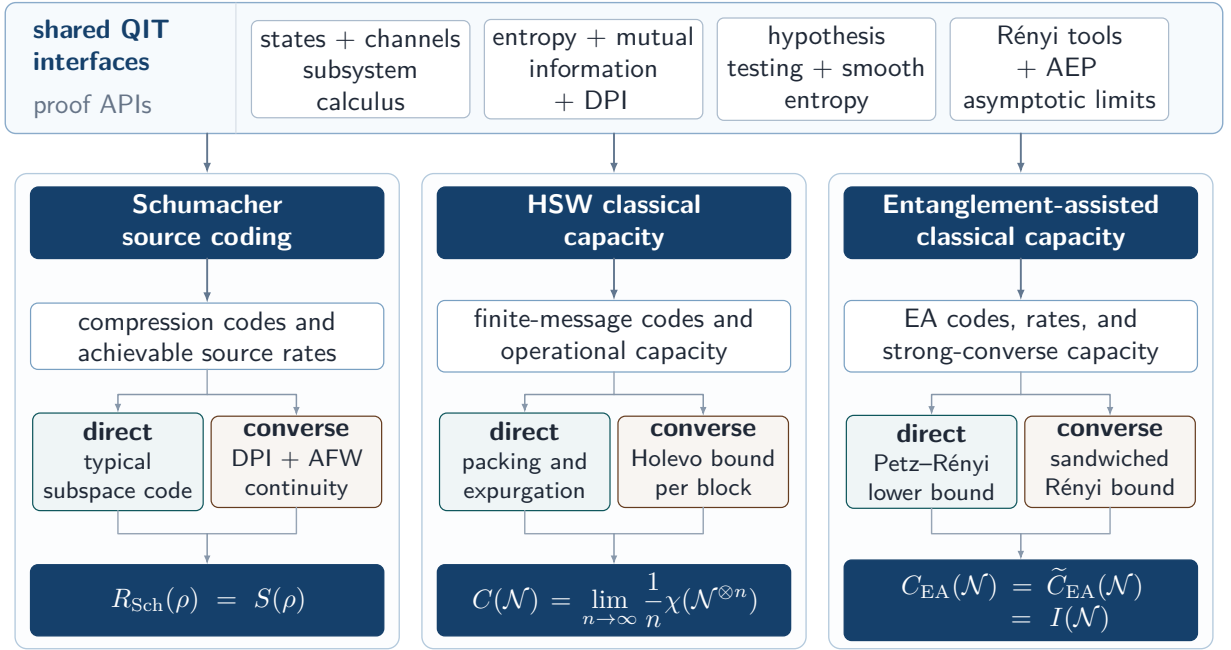
\begin{figure}[t]
    \centering
    \begin{tikzpicture}[
x=1cm,
y=1cm,
every node/.style={font=\normalsize\sffamily,text=QSTInk,outer sep=0pt},
lane/.style={draw=QSTPanelBlueFrame!46,fill=QSTPanelBlueSoft!17,rounded corners=4.5pt,line width=0.50pt,minimum width=5.08cm,minimum height=6.28cm},
sharedtitle/.style={font=\normalsize\sffamily\bfseries,text=QSTPanelBlueDark,align=left,text width=2.62cm,inner sep=0pt},
sharedsub/.style={font=\normalsize\sffamily,text=QSTMuted,align=left,text width=2.62cm,inner sep=0pt},
sharedmodule/.style={draw=QSTFrame!72,fill=white,rounded corners=2.5pt,line width=0.46pt,text width=2.72cm,minimum height=0.92cm,align=center,inner xsep=2.4pt,inner ysep=2.8pt,font=\normalsize\sffamily},
linehead/.style={draw=QSTPanelBlueDark,fill=QSTPanelBlueDark,text=white,rounded corners=3pt,line width=0.50pt,text width=4.46cm,minimum height=0.88cm,align=center,inner xsep=3pt,inner ysep=3pt,font=\normalsize\sffamily\bfseries},
api/.style={draw=QSTPanelBlueFrame!82,fill=white,rounded corners=2.8pt,line width=0.46pt,text width=4.46cm,minimum height=0.82cm,align=center,inner xsep=3pt,inner ysep=3pt,font=\normalsize\sffamily},
direct/.style={draw=QSTDone!72!black,fill=QSTDone!7,rounded corners=2.8pt,line width=0.46pt,text width=2.10cm,minimum height=1.18cm,align=center,inner xsep=2.2pt,inner ysep=3pt,font=\small\sffamily},
converse/.style={draw=QSTWarn!72!black,fill=QSTWarn!6,rounded corners=2.8pt,line width=0.46pt,text width=2.10cm,minimum height=1.18cm,align=center,inner xsep=2.2pt,inner ysep=3pt,font=\small\sffamily},
endpoint/.style={draw=QSTPanelBlueDark,fill=QSTPanelBlueDark,rounded corners=2.8pt,line width=0.50pt,text=white,text width=4.46cm,minimum height=0.94cm,align=center,inner xsep=3pt,inner ysep=3pt,font=\normalsize\sffamily\bfseries},
layerflow/.style={-{Latex[length=1.42mm,width=1.04mm]},line width=0.72pt,draw=QSTNavy!68,line cap=round,line join=round,shorten <=0pt,shorten >=0pt},
branchjoin/.style={line width=0.54pt,draw=QSTNavy!45,line cap=round,line join=round,shorten <=0pt,shorten >=0pt},
branchflow/.style={-{Latex[length=1.22mm,width=0.90mm]},line width=0.56pt,draw=QSTNavy!52,line cap=round,line join=round,shorten <=0pt,shorten >=0pt}
]

\path[use as bounding box] (-8.20,-0.12) rectangle (8.20,8.72);

\draw[draw=QSTPanelBlueFrame!72,fill=QSTPanelBlueSoft!54,line width=0.52pt,rounded corners=4pt] (-8.05,6.92) rectangle (8.05,8.65);
\draw[draw=QSTFrame!34,line width=0.46pt] (-5.00,6.92) -- (-5.00,8.65);

\node[sharedtitle,anchor=north west] at (-7.68,8.42) {shared QIT\\interfaces};
\node[sharedsub,anchor=south west] at (-7.68,7.12) {proof APIs};

\node[sharedmodule] at (-3.35,7.77) {states + channels\\subsystem\\calculus};
\node[sharedmodule] at (-0.27,7.77) {entropy + mutual\\information\\+ DPI};
\node[sharedmodule] at (2.81,7.77) {hypothesis\\testing + smooth\\entropy};
\node[sharedmodule] at (5.89,7.77) {R\'enyi tools\\+ AEP\\asymptotic limits};

\node[lane] (schlane) at (-5.38,3.25) {};
\node[lane] (hswlane) at (0,3.25) {};
\node[lane] (ealane) at (5.38,3.25) {};

\node[linehead] (sch) at (-5.38,5.74) {Schumacher\\source coding};
\node[linehead] (hsw) at (0,5.74) {HSW classical\\capacity};
\node[linehead] (ea) at (5.38,5.74) {Entanglement-assisted\\classical capacity};

\node[api] (schapi) at (-5.38,4.24) {compression codes and\\achievable source rates};
\node[api] (hswapi) at (0,4.24) {finite-message codes and\\operational capacity};
\node[api] (eaapi) at (5.38,4.24) {EA codes, rates, and\\strong-converse capacity};

\node[direct] (schdir) at (-6.56,2.58) {{\normalsize\textbf{direct}}\\typical\\subspace code};
\node[converse] (schconv) at (-4.20,2.58) {{\normalsize\textbf{converse}}\\DPI + AFW\\continuity};

\node[direct] (hswdir) at (-1.18,2.58) {{\normalsize\textbf{direct}}\\packing and\\expurgation};
\node[converse] (hswconv) at (1.18,2.58) {{\normalsize\textbf{converse}}\\Holevo bound\\per block};

\node[direct] (eadir) at (4.20,2.58) {{\normalsize\textbf{direct}}\\Petz--R\'enyi\\lower bound};
\node[converse] (eaconv) at (6.56,2.58) {{\normalsize\textbf{converse}}\\sandwiched\\R\'enyi bound};

\node[endpoint] (schend) at (-5.38,0.75) {$R_{\mathrm{Sch}}(\rho)=S(\rho)$};
\node[endpoint] (hswend) at (0,0.75) {$C(\cN)=\displaystyle\lim_{n\to\infty}\frac{1}{n}\chi(\cN^{\otimes n})$};
\node[endpoint] (eaend) at (5.38,0.75) {$C_{\operatorname{EA}}(\cN)=\widetilde C_{\operatorname{EA}}(\cN)$\\$\quad \quad \; =I(\cN)$};

\coordinate (schSplit) at (-5.38,3.47);
\coordinate (hswSplit) at (0,3.47);
\coordinate (eaSplit) at (5.38,3.47);

\coordinate (schMerge) at (-5.38,1.63);
\coordinate (hswMerge) at (0,1.63);
\coordinate (eaMerge) at (5.38,1.63);

\draw[layerflow] (-5.38,6.92) -- (schlane.north);
\draw[layerflow] (0,6.92) -- (hswlane.north);
\draw[layerflow] (5.38,6.92) -- (ealane.north);

\draw[layerflow] (sch.south) -- (schapi.north);
\draw[layerflow] (hsw.south) -- (hswapi.north);
\draw[layerflow] (ea.south) -- (eaapi.north);

\draw[branchjoin] (schapi.south) -- (schSplit);
\draw[branchjoin] (hswapi.south) -- (hswSplit);
\draw[branchjoin] (eaapi.south) -- (eaSplit);

\draw[branchflow] (schSplit) -| (schdir.north);
\draw[branchflow] (schSplit) -| (schconv.north);
\draw[branchflow] (hswSplit) -| (hswdir.north);
\draw[branchflow] (hswSplit) -| (hswconv.north);
\draw[branchflow] (eaSplit) -| (eadir.north);
\draw[branchflow] (eaSplit) -| (eaconv.north);

\draw[branchjoin] (schdir.south) |- (schMerge);
\draw[branchjoin] (schconv.south) |- (schMerge);
\draw[branchjoin] (hswdir.south) |- (hswMerge);
\draw[branchjoin] (hswconv.south) |- (hswMerge);
\draw[branchjoin] (eadir.south) |- (eaMerge);
\draw[branchjoin] (eaconv.south) |- (eaMerge);

\draw[branchflow] (schMerge) -- (schend.north);
\draw[branchflow] (hswMerge) -- (hswend.north);
\draw[branchflow] (eaMerge) -- (eaend.north);

\end{tikzpicture}
    \caption{Architecture of the coding spine developed in
        Section~\ref{sec:coding-spine}.  Three independent theorem lanes consume
        the reusable finite-dimensional QIT interfaces of
        Section~\ref{sec:building-blocks}.  Within each lane, an operational API
        branches into direct and converse obligations and reassembles at a proved
        asymptotic endpoint.}
    \label{fig:coding-spine-architecture}
\end{figure}

The coding layer presents \LeanQIT{} through the operational progression of
quantum Shannon theory.  We illustrate three coding theorems in important operational tasks, i.e., Schumacher source coding, unassisted noisy-channel classical communication, and noisy-channel classical communication with entanglement
assistance. Figure~\ref{fig:coding-spine-architecture} summarizes the proof architecture that the code listings and endpoint discussions instantiate.

Each subsection follows the same source-to-Lean pattern.  We first state the operational definition and the information-theoretic benchmark in mathematical form, then identify the corresponding Lean API, and finally explain which
direct, converse, and limit obligations are isolated as named declarations.

\subsection{Schumacher Source Coding}
\label{sec:source-coding}

Schumacher compression is the source-coding starting point of QST. Given many copies of a quantum source $\rho$, the theorem identifies the asymptotic
compression rate with the von Neumann entropy~\cite{Schumacher1995}.  The current \LeanQIT{} source-coding layer separates
this theorem into four parts: a compression-code interface, an
unconditional direct coding theorem, a converse theorem, and the final
infimum-rate equality.  This split is useful both mathematically and
editorially.  The direct theorem exposes the typical-subspace construction,
the converse exposes the mutual-information and continuity chain, and the
limit theorem records the operational equality.

For a block compression code $C_n$ acting on $n$ copies of a source state $\rho$, let $W_n$ be the compressed quantum register and define
\begin{equation}
    r_n(C_n)\coloneqq\frac{1}{n}\log \dim W_n .
\end{equation}
A rate $R\in\RR$ is achievable for the source $\rho$ if for all $\delta, \ve>0$,
\begin{equation}
    \begin{aligned}
        \exists N,\ \forall n\ge N,
        \exists C_n:\quad
        r_n(C_n)\le R+\delta,\qquad
        e_{\mathrm{jp}}(C_n,\rho^{\otimes n})\le \ve ,
    \end{aligned}
\end{equation}
where $e_{\mathrm{jp}}$ denotes the joint purification error used by the
Schumacher code interface.

\begin{definition}[Schumacher compression rate]
    \label{def:schumacher-compression-rate}
    The operational Schumacher compression rate of a source $\rho$ is defined by
    \begin{equation}
        R_{\mathrm{Sch}}(\rho) \coloneqq \inf\{R\in\RR: R\ \mathrm{is~achievable~for~ }\rho\}.
    \end{equation}
\end{definition}

\begin{theorem}[Schumacher source coding~\cite{Schumacher1995}]
    \label{thm:schumacher-source-coding}
    For every finite-dimensional source state $\rho$, the operational
    compression rate is equal to the von Neumann entropy $R_{\mathrm{Sch}}(\rho)=-\tr(\rho\log\rho)$.
\end{theorem}

The first Lean interface is deliberately operational.  The predicate
\texttt{IsAchievableSchumacher\allowbreak Rate} records exactly the eventual
blocklength quantifiers in the asymptotic condition above.  The following
noncomputable definition then turns the mathematical infimum of
Definition~\ref{def:schumacher-compression-rate} into Lean's
order-theoretic \texttt{sInf}.
	
\begin{qitleancode}[title={Operational Schumacher rate},
qit focus={\qitHighlightSpan{4}{8}\qitHighlightSpan{10}{11}}]
namespace State

def IsAchievableSchumacherRate (\rho : State a) (R : \mathbbR) : Prop :=
  \forall \delta : \mathbbR, 0 < \delta \to \forall \epsilon : \mathbbR, 0 < \epsilon \to
    \exists N : \mathbbN, \forall n : \mathbbN, n \ge N \to
      \exists (W : Type u), \exists (_ : Fintype W), \exists (_ : DecidableEq W),
        \exists C : SchumacherCompressionCode \rho n W,
          C.rate \le R + \delta \wedge C.jointError \le \epsilon

noncomputable def schumacherCompressionRate (\rho : State a) : \mathbbR :=
  sInf {R : \mathbbR | \rho.IsAchievableSchumacherRate R}

end State
\end{qitleancode}
	
The theorem code is then much smaller because the direct and converse parts
have already been named.  The first theorem says that the entropy rate is the
least achievable rate. The final public endpoint converts that least-element
statement into the infimum equality of
Theorem~\ref{thm:schumacher-source-coding}.
	
\begin{qitleancode}[title={Schumacher compression endpoint},
qit focus={\qitHighlightSpan{3}{6}\qitHighlightSpan{8}{9}}]
namespace State

theorem schumacherRate_isLeast_achievableRates (\rho : State a) :
    IsLeast {R : \mathbbR | \rho.IsAchievableSchumacherRate R} \rho.schumacherRate :=
  \langle schumacher_direct_achievable \rho,
    fun _R hR => schumacher_converse \rho _ hR\rangle

theorem schumacher_data_compression_limit (\rho : State a) :
    \rho.schumacherCompressionRate = \rho.schumacherRate := by
  set S : Set \mathbbR := {R | \rho.IsAchievableSchumacherRate R} with hSdef
  have hleast := \rho.schumacherRate_isLeast_achievableRates
  have hne : S.Nonempty := \langle \rho.schumacherRate, hleast.1\rangle
  have hbdd : BddBelow S := \langle \rho.schumacherRate, hleast.2\rangle
  have h1 : sInf S \le \rho.schumacherRate := csInf_le hbdd hleast.1
  have h2 : \rho.schumacherRate \le sInf S := le_csInf hne hleast.2
  exact le_antisymm h1 h2

end State
\end{qitleancode}
	
In our Lean formalization, the direct theorem
\texttt{State.schumacher\_direct\_achievable}~\cite{Schumacher1995,Wilde2017} constructs the typical-subspace code and bounds its joint purification trace-distance error.  The converse theorem \texttt{State.schumacher\_converse} follows Wilde's five-step route~\cite{Wilde2017}.  The mutual-information dimension bound~\cite{Wilde2017} is exposed as
\texttt{mutualInformation\_le\_two\_log\_card\_right}.  The local
data-processing inequality~\cite{Wilde2017} is supplied by
\texttt{mutualInformation\_dataProcessing\_local\_channels\_ge}.
Alicki--Fannes--Winter continuity of conditional entropy~\cite{AlickiFannes2004,Winter2016EntropyContinuity} is implemented by
\texttt{State.alickiFannesWinter\_conditionalEntropy}.  Tensor additivity of
mutual information~\cite{Wilde2017} is supplied by
\texttt{mutualInformation\_tensorPowerBipartite}.  Finally, the pure-state
marginal entropy identity~\cite{Wilde2017} is assembled from
\texttt{pureVector\_vonNeumann\_eq\_zero} and \\
\texttt{pureVector\_marginalA\_vonNeumann\_eq\_marginalB}.  The limit theorem
\texttt{State.schumacher\_data\_compression\_limit}\\\cite{Schumacher1995,Wilde2017} then turns the direct and converse
inequalities into the infimum equality.

This decomposition illustrates the library's convention where the public theorem is compact, while each proof-theoretic responsibility has a named Lean endpoint.  Several of these endpoints are not specific to source coding.  The
mutual-information dimension bound is reused in other converse arguments, the
local-channel data-processing interface is the same one described in
Section~\ref{subsec:information-dpi}, and the AFW endpoint applies whenever a
trace-distance reliability guarantee must be converted into an entropy bound.

\subsection{Classical Capacity}
\label{sec:hsw}

The second line moves from source compression to noisy-channel coding.  For
unassisted classical communication over a quantum channel $\cN$, the HSW
theorem identifies the operational classical capacity with the regularized
Holevo expression~\cite{Holevo1998,Schumacher1996}.  This theorem is a
demanding formalization target because the final equality depends on several distinct proof layers: finite-block message codes, Holevo information, Hayashi--Nagaoka decoding estimates, packing and expurgation, converse bounds, superadditivity, and the limiting regularized expression.

\LeanQIT{} treats HSW as a completed public capacity endpoint.  The
classical-capacity layer includes the capacity definition, the relevant
Holevo quantities, the Holevo dimension bound, the Hayashi--Nagaoka operator inequality, a derandomized packing lemma, and the final HSW theorem~\cite{HayashiNagaoka2003,KhatriLamiWilde2025Principles}. The library first defines what it means for a real number $R$ to be an achievable classical communication rate. Only after this predicate is in
place does the library define the operational capacity as the supremum of
achievable rates.  The Holevo expression is then introduced as a separate
information quantity, and the HSW theorem proves that the two coincide.

The formal source statements used below follow the Khatri--Lami--Wilde chapter~\cite{KhatriLamiWilde2025Principles} tracked by the Lean proof, with the original direct-coding theorem due to Holevo--Schumacher--Westmoreland and the regularized-capacity formulation presented in modern QST references~\cite{Holevo1998,Schumacher1996,HayashiNagaoka2003}.

\begin{definition}[$(n,|\mathcal M|,\ve)$ classical communication protocol]
    \label{def:hsw-classical-protocol}
    A classical communication protocol over $n$ uses of a quantum channel
    $\cN_{A\to B}$ is specified by the triple $(\mathcal M,\mathcal E_{M\to A^n},\mathcal D_{B^n\to \widehat M})$ where $\mathcal M$ is a finite message set, $M$ is the classical message register with alphabet $\mathcal M$, $\widehat M$ is the decoded message register, $\mathcal E$ is an encoding channel, and $\mathcal D$ is a decoding channel.  For $\ve\in[0,1]$, the protocol is an $(n,|\mathcal M|,\ve)$ classical communication protocol if its maximal error
    probability satisfies
    \begin{equation}
        p_{\mathrm{err}}^*(\mathcal E,\mathcal D;\cN^{\otimes n})\le \ve .
    \end{equation}
    The rate of the protocol is defined by
    \begin{equation}
        R(n,|\mathcal M|)
        \coloneqq
        \frac{1}{n}\log|\mathcal M|.
    \end{equation}
\end{definition}

\begin{definition}[Achievable rate for classical communication]
    \label{def:hsw-achievable-rate}
    A rate $R\in\mathbb R_{\ge 0}$ is an achievable rate for classical
    communication over a quantum channel $\cN$ if, for every
    $\ve\in(0,1]$, every $\delta>0$, and all sufficiently large $n$, there
    exists an $(n,2^{n(R-\delta)},\ve)$ classical communication protocol.
\end{definition}

The Lean predicate represents this source shorthand by an explicit finite
message set $\mathcal M_n$ satisfying
\begin{equation}
    \frac{1}{n}\log|\mathcal M_n|\ge R-\delta,
\end{equation}
which is the kernel-level form of the same rate condition.

\begin{definition}[Classical capacity]
    \label{def:hsw-classical-capacity}
    The classical capacity of a quantum channel $\cN$ is defined as the
    supremum of its achievable classical communication rates:
    \begin{equation}
        C(\cN)
        \coloneqq
        \sup\{R:R\mathrm{~is~an~achievable~rate~for~}\cN\}.
    \end{equation}
\end{definition}

\begin{definition}[Holevo information]
    \label{def:hsw-holevo-information}
    For a finite-dimensional quantum channel $\cN$, the Holevo information
    is defined by
    \begin{equation}
            \chi(\cN)
            \coloneqq
            \sup_{\{p_x,\rho_x\}_{x\in\mathcal X}}
            \left[
            S\!\left(\sum_x p_x\,\cN(\rho_x)\right)
            -\sum_x p_x S(\cN(\rho_x))
            \right],
    \end{equation}
    where the supremum is over finite ensembles of input states with
    $p_x\ge0$ and $\sum_xp_x=1$.
\end{definition}

\begin{theorem}[Classical capacity of a quantum channel~\cite{Holevo1998,Schumacher1996}]
    \label{thm:hsw-capacity}
    For every finite-dimensional quantum channel $\cN$, the classical capacity
    is equal to the regularized Holevo expression
    \begin{equation}
        C(\cN)
        =
        \lim_{n\to\infty}\frac{1}{n}\chi(\cN^{\otimes n}).
    \end{equation}
\end{theorem}

The Lean formalization follows the same separation as the mathematics.  The
first code block contains only the operational side of
Definitions~\ref{def:hsw-classical-protocol}--\ref{def:hsw-classical-capacity}:
achievability, upper bounds on achievable rates, and the supremum defining
capacity.
	
\begin{qitleancode}[title={Operational classical capacity},
qit focus={\qitHighlightSpan{4}{5}\qitHighlightSpan{8}{9}\qitHighlightSpan{14}{15}}]
namespace Channel

def IsAchievableClassicalRate (R : \mathbbR) : Prop :=
  \forall \delta : \mathbbR, 0 < \delta \to \forall \epsilon : \mathbbR, 0 < \epsilon \to
    \exists N0 : \mathbbN, \forall n : \mathbbN, n \ge N0 \to
      \exists (M : Type u), \exists (_ : Fintype M),
        \exists (_ : DecidableEq M), \exists (_ : Nonempty M),
          \exists C : HSWClassicalCode N n M,
            C.rate \ge R - \delta \wedge C.maxErrorAtMost \epsilon

def IsClassicalRateUpperBound (B : \mathbbR) : Prop :=
  \forall R : \mathbbR, N.IsAchievableClassicalRate R \to R \le B

def classicalCapacity : \mathbbR :=
  sSup {R : \mathbbR | N.IsAchievableClassicalRate R}

end Channel
\end{qitleancode}
	
The Holevo side is a different interface. The operational definition of $C(\cN)$ does not mention Holevo information, and the theorem precisely states that the operational supremum equals the regularized Holevo expression.

\begin{qitleancode}[title={Regularized Holevo information}]
namespace Channel

def blockHolevoInformation (n : \mathbbN) : \mathbbR :=
  (N.tensorPower n).holevoInformation

def regularizedHolevoRateValues : Set \mathbbR :=
  {R : \mathbbR | \exists n : \mathbbN, 0 < n \wedge
    R = N.blockHolevoInformation n / (n : \mathbbR)}

def regularizedHolevoInformation : \mathbbR :=
  sSup N.regularizedHolevoRateValues

end Channel
\end{qitleancode}
	
	Between these two sides, the formal proof contains finite-block bridge
	declarations that are easy to miss in the final theorem statement.  The
	packing lemma first produces a small average-error decoder from typical
	projectors and the Hayashi--Nagaoka operator inequality~\cite{HayashiNagaoka2003,KhatriLamiWilde2025Principles}.  A separate expurgation
	layer turns average error into maximal error, paying the standard
	one-bit-per-block rate loss~\cite{KhatriLamiWilde2025Principles}.  Lean records
	these transitions as named witnesses consumed by
	\texttt{hsw\_regularizedHolevoInformation\_direct}.  This is the passage from
	analytic estimates on output states to an actual finite message code, and it
	is the kind of interface needed again in future coding theorems with different
	ensembles, decoders, or error criteria.
	
	Our Lean formulation names the converse upper bound and the direct achievability family separately, uses them to establish nonemptiness and boundedness of the achievable-rate set, and then proves the two inequalities required by the operational
	supremum.
	
\begin{qitleancode}[title={HSW capacity squeeze},
qit focus={\qitHighlightSpan{5}{11}}]
namespace Channel

theorem classicalCapacity_eq_regularizedHolevoInformation
    [Nonempty a] [Nonempty b] (N : Channel a b) :
    N.classicalCapacity = N.regularizedHolevoInformation := by
  have hconv : N.IsClassicalRateUpperBound N.regularizedHolevoInformation :=
    hsw_regularizedHolevoInformation_converse N
  have hdirect :
      \forall R : \mathbbR, R < N.regularizedHolevoInformation \to
        N.IsAchievableClassicalRate R :=
    hsw_regularizedHolevoInformation_direct N
  have hnonempty :
      ({R : \mathbbR | N.IsAchievableClassicalRate R} : Set \mathbbR).Nonempty := by
    refine \langle N.regularizedHolevoInformation - 1, ?_\rangle
    exact hdirect (N.regularizedHolevoInformation - 1) (by linarith)
  have hbounded : BddAbove {R : \mathbbR | N.IsAchievableClassicalRate R} := by
    exact \langle N.regularizedHolevoInformation, fun R hR => hconv R hR\rangle
  apply le_antisymm
  \cdot unfold classicalCapacity
    exact csSup_le hnonempty fun R hR => hconv R hR
  \cdot rw [le_iff_forall_pos_lt_add]
    intro \eta h\eta
    have hAch :
        N.IsAchievableClassicalRate
          (N.regularizedHolevoInformation - \eta / 2) :=
      hdirect (N.regularizedHolevoInformation - \eta / 2) (by linarith)
    have hle :
        N.regularizedHolevoInformation - \eta / 2 \le N.classicalCapacity := by
      unfold classicalCapacity
      exact le_csSup hbounded hAch
    linarith

end Channel
\end{qitleancode}
	
The first branch applies the converse to every achievable rate, proving that
\texttt{sSup} is at most the regularized Holevo information.  The second uses
direct achievability at
$\operatorname{regularizedHolevoInformation}(\cN)-\eta/2$ and the defining
property of \texttt{sSup} to obtain the reverse inequality.  The thin public
wrapper that pairs this equality with the epsilon-limit statement is retained
for API discovery in Appendix~\ref{app:endpoint-wrappers}.

\subsection{Entanglement-assisted classical capacity}
\label{sec:ea}

The third line keeps the noisy-channel setting but changes the communication
resource.  With unlimited shared entanglement, classical communication over a
quantum channel is governed by a single-letter channel mutual information
rather than by a regularized Holevo expression~\cite{Bennett1999,Bennett2002}.  The presentation below follows the modern
Khatri--Lami--Wilde organization of the entanglement-assisted classical capacity
theorem~\cite{KhatriLamiWilde2025Principles}. We define the one-shot protocol and error criterion, define asymptotic achievable and strong-converse rates, and
only then assemble the capacity identity.  This ordering matters for
formalization because the final equality is not the primitive object in Lean. It is assembled from typed operational definitions and separately named direct and converse inputs.

The source proof route itself combines several strands of QST.  The
one-shot viewpoint for entanglement-assisted communication originates in the
one-shot EA literature of Datta--Hsieh~\cite{DattaHsieh2011}.  The
Khatri--Lami--Wilde direct branch then
uses position-based coding, sequential decoding, and the quantum union bound
to obtain a hypothesis-testing lower bound~\cite{DattaHsieh2011,AnshuJainWarsi2019,QiWangWilde2018,OskoueiManciniWilde2019,KhatriLamiWilde2025Principles}.
In this organization, position-based coding is traced to
Anshu--Jain--Warsi, the EA one-shot lower bound to Qi--Wang--Wilde, and the
quantum-union-bound proof route to Oskouei--Mancini--Wilde~\cite{AnshuJainWarsi2019,QiWangWilde2018,OskoueiManciniWilde2019,KhatriLamiWilde2025Principles}.
The Petz--R\'enyi lower bound used in the asymptotic direct proof is obtained
from this hypothesis-testing lower bound by comparison, then tensorized and
sent to the $\alpha\to1^{-}$ limit~\cite{QiWangWilde2018,KhatriLamiWilde2025Principles}.  The converse branch uses the
Matthews--Wehner finite-blocklength meta-converse through
hypothesis-testing mutual information, and the strong-converse upgrade uses
the Gupta--Wilde sandwiched-R\'enyi and completely bounded norm route~\cite{MatthewsWehner2012,GuptaWilde2013,KhatriLamiWilde2025Principles}.
\LeanQIT{} preserves this decomposition: the lower, weak-converse, and
strong-converse proof obligations are exposed as named theorems before the
final strong-converse endpoint is stated.
	
\begin{definition}[One-shot entanglement-assisted classical capacity]
    \label{def:ea-one-shot-capacity}
    An $(|\mathcal M|,\ve)$ entanglement-assisted classical communication
    protocol for a channel $\cN_{A\to B}$ consists of a message set
    $\mathcal M$, a shared state $\Psi_{A'B'}$, an encoding channel
    $\mathcal E_{M'A'\to A}$, and a decoding channel
    $\mathcal D_{BB'\to \widehat M}$, with maximal error
    $p_{\mathrm{err}}^*((\Psi,\mathcal E,\mathcal D);\cN)\le \ve$.  The
    registers $A'$ and $B'$ are the sender's and receiver's shares of the
    prior entanglement, $M'$ is the classical message register before
    encoding, and $\widehat M$ is the decoded message register.  The
    one-shot $\ve$-error entanglement-assisted classical capacity is defined by
    \begin{equation}
        \begin{aligned}
            C_{\operatorname{EA}}^{\ve}(\cN)
            &\coloneqq
            \sup_{(\mathcal M,\Psi,\mathcal E,\mathcal D)}
            \bigl\{\log|\mathcal M|:
            p_{\mathrm{err}}^*((\Psi,\mathcal E,\mathcal D);\cN)\le \ve\bigr\}.
        \end{aligned}
    \end{equation}
\end{definition}

For the asymptotic formulation, an $(n,|\mathcal M|,\ve)$
entanglement-assisted protocol for $\cN_{A\to B}$ is an
$(|\mathcal M|,\ve)$ protocol for $\cN^{\otimes n}$.  Its rate is defined by
\begin{equation}
    R(n,|\mathcal M|)
    \coloneqq
    \frac{1}{n}\log|\mathcal M|.
\end{equation}
A real number $R$ is an achievable entanglement-assisted classical rate if
for every $\ve\in(0,1]$ and every $\delta>0$, all sufficiently large
$n$ admit an $(n,|\mathcal M_n|,\ve)$ protocol satisfying
\begin{equation}
    \frac{1}{n}\log|\mathcal M_n|\ge R-\delta .
\end{equation}

\begin{definition}[Entanglement-assisted capacity and strong-converse capacity]
    \label{def:ea-capacities}
    The entanglement-assisted classical capacity is defined as the supremum of
    achievable rates:
    \begin{equation}
        C_{\operatorname{EA}}(\cN)
        \coloneqq
        \sup\{R:R\ \mathrm{is~an~achievable~EA~classical~rate~for~}\cN\}.
    \end{equation}
    A real number $R$ is a strong-converse EA classical rate if for every
    $\ve\in[0,1)$ and every $\delta>0$, all sufficiently large $n$ satisfy:
    every $(n,|\mathcal M_n|,\ve)$ protocol has
    \begin{equation}
        \frac{1}{n}\log|\mathcal M_n| < R+\delta .
    \end{equation}
    The strong-converse entanglement-assisted classical capacity is defined by
    \begin{equation}
        \widetilde C_{\operatorname{EA}}(\cN)
        \coloneqq
        \inf\{R:R\ \mathrm{is~a~strong-converse~EA~classical~rate~for~}\cN\}.
    \end{equation}
\end{definition}
	
\begin{definition}[Channel mutual information]
    \label{def:ea-channel-mutual-information}
    For a channel $\cN_{A\to B}$, the channel mutual information is defined by
    \begin{equation}
        \begin{aligned}
            I(\cN)
            &\coloneqq
            \sup_{\psi_{RA}} I(R;B)_{\omega},
            \qquad
            \omega_{RB}=(\id_R\otimes \cN_{A\to B})(\psi_{RA}),
        \end{aligned}
    \end{equation}
    where the supremum is over pure input-reference states with $R$ isomorphic to $A$, and the state mutual information is defined by
    \begin{equation}
        I(R;B)_\omega=H(R)_\omega+H(B)_\omega-H(RB)_\omega=D(\omega_{RB}\Vert \omega_R\otimes\omega_B).
    \end{equation}
\end{definition}

\begin{theorem}[Entanglement-assisted classical capacity~\cite{Bennett1999,Bennett2002,GuptaWilde2013}]
    \label{thm:ea-capacity-kw}
    For every finite-dimensional quantum channel $\cN$, the entanglement-assisted
    classical capacity and the strong-converse entanglement-assisted classical
    capacity are both equal to the channel mutual information:
    \begin{equation}
        C_{\operatorname{EA}}(\cN)
        =
        \widetilde C_{\operatorname{EA}}(\cN)
        =
        I(\cN).
    \end{equation}
\end{theorem}
	
Theorem~\ref{thm:ea-capacity-kw} is the Bennett--Shor--Smolin--Thapliyal
(BSST) capacity theorem together with the modern strong-converse formulation~\cite{Bennett1999,Bennett2002,GuptaWilde2013,KhatriLamiWilde2025Principles}. The Lean API is organized to match Definitions~\ref{def:ea-one-shot-capacity}--\ref{def:ea-channel-mutual-information} and Theorem~\ref{thm:ea-capacity-kw}, but the formal proof is easiest to read through the same three source-to-Lean transitions used in the Khatri--Lami--Wilde organization~\cite{KhatriLamiWilde2025Principles}.  First, a
source one-shot direct proof is exposed as a Lean direct-achievability
theorem.  Second, a source one-shot converse proof is exposed as a Lean
one-shot upper-bound theorem.  Third, the asymptotic Khatri--Lami--Wilde assembly
turns the direct and converse inputs into the Lean final capacity theorem~\cite{Bennett1999,Bennett2002,KhatriLamiWilde2025Principles}.  The operational
definitions are displayed first because all three transitions consume the
same typed coding interface.

\paragraph{Shared operational definitions.}
The shared interface fixes the finite message type, the two entanglement
registers, the maximal-error predicate, the asymptotic rate quantifiers, and
the supremum defining $C_{\operatorname{EA}}(\cN)$.  These operational
objects are consumed later by both the one-shot achievability theorem and the
one-shot converse theorem; the information formula is deliberately not built
into the coding definition.
	
\begin{qitleancode}[title={Operational EA capacity},
qit focus={\qitHighlightSpan{9}{10}\qitHighlightSpan{13}{14}\qitHighlightSpan{19}{20}\qitHighlightSpan{22}{23}}]
namespace Channel

def oneShotEntanglementAssistedClassicalCapacity (\epsilon : \mathbbR) : \mathbbR :=
  sSup {R : \mathbbR |
    \exists (M : Type u), \exists (_ : Fintype M),
      \exists (_ : DecidableEq M), \exists (_ : Nonempty M),
        \exists (EA : Type u), \exists (_ : Fintype EA), \exists (_ : DecidableEq EA),
          \exists (EB : Type u), \exists (_ : Fintype EB), \exists (_ : DecidableEq EB),
            \exists C : EntanglementAssistedClassicalCode N 1 M EA EB,
              C.maxErrorAtMost \epsilon \wedge R = C.rate}

def IsAchievableEntanglementAssistedClassicalRate (R : \mathbbR) : Prop :=
  \forall \delta : \mathbbR, 0 < \delta \to \forall \epsilon : \mathbbR, 0 < \epsilon \to
    \exists N0 : \mathbbN, \forall n : \mathbbN, n \ge N0 \to
      \exists (M : Type u), \exists (_ : Fintype M),
        \exists (_ : DecidableEq M), \exists (_ : Nonempty M),
          \exists (EA : Type u), \exists (_ : Fintype EA), \exists (_ : DecidableEq EA),
            \exists (EB : Type u), \exists (_ : Fintype EB), \exists (_ : DecidableEq EB),
              \exists C : EntanglementAssistedClassicalCode N n M EA EB,
                C.rate \ge R - \delta \wedge C.maxErrorAtMost \epsilon

def entanglementAssistedClassicalCapacity : \mathbbR :=
  sSup {R : \mathbbR | N.IsAchievableEntanglementAssistedClassicalRate R}

end Channel
\end{qitleancode}
	
	The same definition layer also records the strong-converse predicate and the
	information quantity that appears in the BSST formula.  The strong-converse
	predicate reverses the coding quantifiers: above rate $R+\delta$, every
	sufficiently long code with error below one must have rate below that
	threshold.  The channel mutual information is a separate supremum over
	purifications, so the library records $I(\cN)$ as an information quantity
	rather than as part of the capacity definition~\cite{Bennett1999,Bennett2002,KhatriLamiWilde2025Principles}.
	
\begin{qitleancode}[title={EA information and strong converse},
qit focus={\qitHighlightSpan{4}{5}\qitHighlightSpan{9}{10}\qitHighlightSpan{16}{17}}]
namespace Channel

def IsStrongConverseEntanglementAssistedClassicalRate (R : \mathbbR) : Prop :=
  \forall \delta : \mathbbR, 0 < \delta \to \forall \epsilon : \mathbbR, 0 \le \epsilon \to \epsilon < 1 \to
    \exists N0 : \mathbbN, \forall n : \mathbbN, n \ge N0 \to
      \forall (M : Type u) [Fintype M] [DecidableEq M] [Nonempty M],
        \forall (EA : Type u) [Fintype EA] [DecidableEq EA],
          \forall (EB : Type u) [Fintype EB] [DecidableEq EB],
            \forall C : EntanglementAssistedClassicalCode N n M EA EB,
              C.maxErrorAtMost \epsilon \to C.rate < R + \delta

def entanglementAssistedInformation : \mathbbR :=
  sSup (Set.range fun \psi : PureVector (Prod a a) =>
    N.entanglementAssistedMutualInformation \psi)

def strongConverseEntanglementAssistedClassicalCapacity : \mathbbR :=
  sInf {R : \mathbbR | N.IsStrongConverseEntanglementAssistedClassicalRate R}

end Channel
\end{qitleancode}
	
\paragraph{Source one-shot direct proof to Lean direct theorem.}
The source direct branch separates a concrete one-shot code witness from the
later supremum argument.  This distinction matters because a real lower bound
need not itself be the logarithm of a finite message cardinality.  The Lean
object below records the finite message set, the actual entanglement-assisted
code, the maximal-error guarantee, and the certified rate inequality.  In the
Khatri--Lami--Wilde direct route, position-based coding and sequential decoding
produce witnesses of this shape~\cite{AnshuJainWarsi2019,QiWangWilde2018,OskoueiManciniWilde2019,KhatriLamiWilde2025Principles}.
Petz--R\'enyi comparison and supremum closure then turn them into the Lean theorem
\texttt{entanglementAssistedInformation\_isAchievable\_of\_oneShotPetzLowerBound}~\cite{DattaHsieh2011,AnshuJainWarsi2019,QiWangWilde2018,OskoueiManciniWilde2019,KhatriLamiWilde2025Principles}.
The barred information quantities in this route fix the side-information
reference to be the actual marginal rather than optimizing over an auxiliary
state.  For a pure input-reference state $\psi_{RA}$, write
$\omega_{RB}=(\id_R\otimes\cN)(\psi_{RA})$.  The barred
hypothesis-testing channel mutual information is defined by
\begin{equation}
    \overline I_{\mathrm H}^{\varepsilon}(\cN)
    \coloneqq
    \sup_{\psi_{RA}}
    D_{\mathrm H}^{\varepsilon}
    \!\left(\omega_{RB}\middle\|\omega_R\otimes\omega_B\right).
\end{equation}
For $\rho\in\mathcal D(\mathcal H)$, $\sigma\succeq0$, and
$\alpha\in(0,1)$, the Petz--R\'enyi divergence is defined by
\begin{equation}
D_\alpha^{\mathrm P}(\rho\Vert\sigma)
\coloneqq \frac{1}{\alpha-1}
\log\operatorname{Tr}(\rho^\alpha\sigma^{1-\alpha})
\end{equation}
when the trace is positive, and set
\(D_\alpha^{\mathrm P}(\rho\Vert\sigma)=+\infty\) when it vanishes.
% For $0<\alpha<1$, the Petz--R\'enyi divergence used by the direct branch
% is defined by
% \begin{equation}
% 	D_{\alpha}^{\mathrm P}(\rho\Vert\sigma)
% 	=
% 	\frac{1}{\alpha-1}
% 	\log\tr\!\left(\rho^{\alpha}\sigma^{1-\alpha}\right),
% \end{equation}
The corresponding channel quantity is defined by
\begin{equation}
    \overline I_{\alpha}^{\mathrm P}(\cN)
    \coloneqq \sup_{\psi_{RA}\ {\rm pure}}
    D_{\alpha}^{\mathrm P}
    \!\left(\omega_{RB}\middle\|\omega_R\otimes\omega_B\right).
\end{equation}
The one-shot source theorem formalized by the library can then be stated as
follows.
	
\begin{theorem}[One-shot EA achievability lower bounds~\cite{DattaHsieh2011,AnshuJainWarsi2019,QiWangWilde2018,OskoueiManciniWilde2019}]
    \label{thm:ea-one-shot-achievability}
    Let $\cN$ be a finite-dimensional quantum channel, let
    $0<\eta<\ve$, and let $0<\alpha<1$.  Then
    \begin{equation}
        \overline I_{\mathrm H}^{\ve-\eta}(\cN)
        -
        \log\!\left(\frac{4\ve}{\eta^2}\right)
        \le
        C_{\operatorname{EA}}^{\ve}(\cN),
    \end{equation}
    and
    \begin{equation}
        \overline I_{\alpha}^{\mathrm P}(\cN)
        -
        \frac{\alpha}{1-\alpha}
        \log\!\left(\frac{1}{\ve-\eta}\right)
        -
        \log\!\left(\frac{4\ve}{\eta^2}\right)
        \le
        C_{\operatorname{EA}}^{\ve}(\cN).
    \end{equation}
\end{theorem}

\begin{qitleancode}[title={One-shot achievability witness}]
namespace QIT

structure EntanglementAssistedOneShotAchievabilityWitness (N : Channel a b)
    (\epsilon lowerBound : \mathbbR)
    (M : Type u) [Fintype M] [DecidableEq M] [Nonempty M]
    (EA : Type u) [Fintype EA] [DecidableEq EA]
    (EB : Type u) [Fintype EB] [DecidableEq EB] where
  code : EntanglementAssistedClassicalCode N 1 M EA EB
  maxError_le : code.maxErrorAtMost \epsilon
  lowerBound_le_rate : lowerBound \le code.rate

end QIT
\end{qitleancode}
	
\paragraph{Source one-shot converse proof to Lean upper-bound theorem.}
The source converse proof uses an extended-real one-shot capacity because
hypothesis-testing quantities may take the value $+\infty$. The Matthews--Wehner meta-converse first bounds a finite-message code by an
optimized hypothesis-testing mutual information, after which the
Khatri--Lami--Wilde route converts that bound into weak-converse and
sandwiched-R\'enyi upper bounds~\cite{MatthewsWehner2012,GuptaWilde2013,KhatriLamiWilde2025Principles}.
Working in $\overline{\mathbb R}$ avoids inserting artificial boundedness
hypotheses at the one-shot level.
	
\begin{qitleancode}[title={Extended-real one-shot EA capacity}]
namespace Channel

def oneShotEntanglementAssistedClassicalCapacityE (\epsilon : \mathbbR) : EReal :=
  sSup {R : EReal |
    \exists (M : Type u), \exists (_ : Fintype M),
      \exists (_ : DecidableEq M), \exists (_ : Nonempty M),
        \exists (EA : Type u), \exists (_ : Fintype EA), \exists (_ : DecidableEq EA),
          \exists (EB : Type u), \exists (_ : Fintype EB), \exists (_ : DecidableEq EB),
            \exists C : EntanglementAssistedClassicalCode N 1 M EA EB,
              C.maxErrorAtMost \epsilon \wedge R = (C.rate : EReal)}

end Channel
\end{qitleancode}
	
The sandwiched-R\'enyi branch uses the notation of the source theorem,
not a Lean identifier promoted to mathematical notation.  For a bipartite
state $\rho_{RB}$, the corresponding source quantity is defined by
\begin{equation}
    \widetilde I_{\alpha}(R;B)_{\rho}
    \coloneqq
    \inf_{\sigma_B}
    \widetilde D_{\alpha}
    \!\left(\rho_{RB}\middle\|\rho_R\otimes\sigma_B\right),
    \qquad
    \widetilde I_{\alpha}(\cN)
    \coloneqq
    \sup_{\psi_{RA}\ {\rm pure}}
    \widetilde I_{\alpha}(R;B)_
    {(\operatorname{id}_{R}\otimes\cN)(\psi_{RA})}.
\end{equation}
The channel quantity $\widetilde I_{\alpha}(\cN)$ is represented in
Lean by the channel-level sandwiched-R\'enyi mutual information.  The two
source right-hand sides are represented by the following weak-converse and
extended-real sandwiched-R\'enyi definitions.
	
	\Needspace{16\baselineskip}
\begin{qitleancode}[title={One-shot converse bounds}]
namespace Channel

def entanglementAssistedWeakConverseBound (\epsilon : \mathbbR) : \mathbbR :=
  (N.entanglementAssistedInformation +
      EntanglementAssistedWeakConverse.binaryEntropy \epsilon) / (1 - \epsilon)

def entanglementAssistedSandwichedOneShotConverseBoundE
    (epsilon alpha : \mathbbR) : EReal :=
  N.sandwichedRenyiMutualInformationE alpha +
    ((alpha / (alpha - 1) * log2 (1 / (1 - epsilon)) : \mathbbR) : EReal)

end Channel
\end{qitleancode}
	
	The source theorem combines the Matthews--Wehner hypothesis-testing
	meta-converse with the Gupta--Wilde sandwiched-R\'enyi strong-converse
	conversion.  We follow the Khatri--Lami--Wilde EA-capacity organization of these
	ingredients~\cite{MatthewsWehner2012,GuptaWilde2013,KhatriLamiWilde2025Principles}.
	
	\begin{theorem}[One-shot EA converse upper bounds~\cite{MatthewsWehner2012,GuptaWilde2013}]
		\label{thm:ea-one-shot-converse}
		Let $\cN$ be a finite-dimensional quantum channel, let
		$0\le\ve<1$, and let $\alpha>1$.  Then
		\begin{equation}
			C_{\operatorname{EA}}^{\ve}(\cN)
			\le
			\frac{1}{1-\ve}[I(\cN)+h_2(\ve)],
		\end{equation}
		and
		\begin{equation}
			C_{\operatorname{EA}}^{\ve}(\cN)\le
			\widetilde I_{\alpha}(\cN)+\frac{\alpha}{\alpha-1}\log\!\left(\frac{1}{1-\ve}\right).
		\end{equation}
	\end{theorem}
	
	The Lean proof route is exposed by named intermediate endpoints rather than by
	the low-level constructor proof of the final package theorem.  Both branches
	first invoke the one-shot hypothesis-testing meta-converse.  The weak branch
	then applies the channel weak-converse comparison.  The code below selects the
	sandwiched branch because its \texttt{.trans} composition makes the proof
	architecture explicit: the same meta-converse is followed by the optimized
	hypothesis-testing-to-sandwiched-R\'enyi comparison.
	
	\Needspace{18\baselineskip}
\begin{qitleancode}[title={Sandwiched-R\'enyi one-shot converse},
qit focus={\qitHighlightSpan{7}{13}}]
namespace Channel

theorem oneShotEntanglementAssistedClassicalCapacityE_le_sandwichedRenyiMutualInformationE_add
    [Nonempty a] {epsilon alpha : \mathbbR}
    (hepsilon_nonneg : 0 \le epsilon) (hepsilon_lt_one : epsilon < 1)
    (halpha : 1 < alpha) :
    N.oneShotEntanglementAssistedClassicalCapacityE epsilon \le
      N.sandwichedRenyiMutualInformationE alpha +
        ((alpha / (alpha - 1) * log2 (1 / (1 - epsilon)) : \mathbbR) : EReal) :=
  (N.oneShotEntanglementAssistedClassicalCapacityE_le_hypothesisTestingMutualInformationE
    hepsilon_nonneg).trans
    (N.hypothesisTestingMutualInformationE_le_sandwichedRenyiMutualInformationE_add
      hepsilon_nonneg hepsilon_lt_one halpha)

end Channel
\end{qitleancode}
	
\paragraph{Source asymptotic Khatri--Lami--Wilde assembly to Lean final theorem.}
The asymptotic Khatri--Lami--Wilde layer is the bridge from one-shot bounds to
capacity~\cite{KhatriLamiWilde2025Principles}.  On the converse side, the
asymptotic upper-input declaration lifts the one-shot estimates to blocklength $n$,
single-letterizes the sandwiched-R\'enyi channel quantity using product-channel
additivity and cb-norm comparison, and then takes the $\alpha\to1^{+}$
endpoint, following the Gupta--Wilde strong-converse route~\cite{GuptaWilde2013,KhatriLamiWilde2025Principles}.  Its conclusion supplies
both the ordinary capacity upper bound and the strong-converse upper bound.
The source-facing declaration that forwards these two asymptotic inputs is a
thin API wrapper and is reproduced in
Appendix~\ref{app:endpoint-wrappers}.

\paragraph{Operational capacity squeeze.}
The final assembly can now focus on the order-theoretic capacity argument.  The
direct side is supplied by the Petz-achievability input, which proves that
$I(\cN)$ is achievable from the one-shot Petz lower bound, tensor-power
lower bounds for the barred Petz--R\'enyi channel quantity, and the
$\alpha\to1^{-}$ Petz endpoint.  The converse side is supplied by the
source converse witness built from the sandwiched-limit upper input.  The
declaration below isolates the ordinary-capacity squeeze: the converse witness
gives the upper inequality, while achievability places $I(\cN)$ inside the set
whose supremum defines the operational capacity.
	
\begin{qitleancode}[title={EA capacity squeeze},
qit focus={\qitHighlightSpan{5}{9}}]
namespace Channel

theorem entanglementAssistedClassicalCapacity_eq_information_of_achievable_of_sourceConverseWitness
    (hach :
      N.IsAchievableEntanglementAssistedClassicalRate
        N.entanglementAssistedInformation)
    (hconv : EntanglementAssistedSourceConverseWitnessFamily N) :
    N.entanglementAssistedClassicalCapacity =
      N.entanglementAssistedInformation := by
  exact le_antisymm
    (N.entanglementAssistedClassicalCapacity_le_information_of_sourceConverseWitness
      hach hconv)
    (by
      unfold entanglementAssistedClassicalCapacity
      exact le_csSup
        \langle N.entanglementAssistedInformation,
          N.entanglementAssisted_information_isUpperBound_of_sourceConverseWitness hconv\rangle
        hach)

end Channel
\end{qitleancode}
	
The strong-converse equality uses the dual \texttt{sInf} interface and the
ordering between ordinary and strong-converse capacities.  The final public
theorem instantiates both abstract squeezes with the Petz achievability witness
and the sandwiched-R\'enyi converse family, rather than reproving either route
inline~\cite{Bennett1999,Bennett2002,GuptaWilde2013,KhatriLamiWilde2025Principles}.
Its short source-facing wrapper is recorded in
Appendix~\ref{app:endpoint-wrappers}.

\paragraph{Formalization obligations exposed by the endpoint.}
The EA endpoint closes the coding spine by making the one-shot and R\'enyi
layers visible inside an asymptotic capacity theorem.  The final proof
depends on one-shot estimates, R\'enyi limits, channel additivity, cb-norm
comparisons, and asymptotic operational lifting.  The named lemmas include
the position-based coding construction, the sequential-decoding quantum union
bound, finite-message witnesses for strict one-shot lower rates, the
Petz--R\'enyi one-shot lower bound that supplies achievability, the
tensor-power and $\alpha\to1^{-}$ Petz limits used in the direct branch,
the hypothesis-testing comparator and decoder data-processing steps used in
the meta-converse, the sandwiched-R\'enyi PSD-reference DPI from
Section~\ref{subsec:information-dpi}, optimized sandwiched-R\'enyi
mutual-information data processing under local channels, and the
product-channel additivity and cb-norm bridges used in the
Gupta--Wilde strong-converse route as organized in the
Khatri--Lami--Wilde presentation~\cite{DattaHsieh2011,AnshuJainWarsi2019,QiWangWilde2018,OskoueiManciniWilde2019,MatthewsWehner2012,GuptaWilde2013,KhatriLamiWilde2025Principles}.
This granularity is the
main scientific value of the formalization: it records not only the capacity
formula, but also which one-shot, R\'enyi, and asymptotic ingredients are
required to certify the formula.

Across the three theorem lines, Section~\ref{sec:coding-spine} illustrates
the central discipline of \LeanQIT{}: operational rates are defined
independently of their information formulae, direct and converse claims are
named before they are assembled, and asymptotic endpoint theorems are exposed
as compact consequences of the shared QIT interface.  The three endpoints
therefore validate the same architectural claim across distinct coding models:
finite-dimensional analytic tools become operational QST only after they are
connected to typed code objects, rate predicates, and explicit direct/converse
assembly.
	
	% ===================================================================
	\section{Conclusion}
	\label{sec:conclusion}
	
This work presents \LeanQIT{}, a reusable finite-dimensional QIT substrate whose compositional design is validated through operational quantum Shannon theory. Its central contribution is a formal interface connecting analytic information quantities to finite-block protocol witnesses and asymptotic coding limits. By keeping operational objects independent of the formulae that later characterize them, the library turns capacity statements into kernel-checkable theorems relating separately defined mathematical structures.

The Schumacher, HSW, and entanglement-assisted theorem lines provide complementary stress tests across source and channel coding, direct and converse reasoning, one-shot and asymptotic methods, and regularized and single-letter formulae. Together, they demonstrate that theorem-scale QST formalization can be organized around reusable interfaces rather than theorem-specific proof scripts. The same substrate also identifies a concrete path toward further coding theorems: quantum and private capacities require coherent-information, privacy, and regularization layers; channel simulation and quantum reverse-Shannon theorems require operational simulation criteria and explicit resource accounting; feedback-assisted tradeoffs and resource inequalities require
rate-region interfaces that track multiple resources simultaneously. These extensions can reuse the present state and channel substrate, information quantities, finite-resource bounds, and asymptotic APIs while exposing only their genuinely new obligations.

The architecture also gives AI-assisted formalization a disciplined point of entry. Its near-term value lies in formal retrieval, proof completion, assumption auditing, and maintenance of theorem chains. The present work does not implement or evaluate an autonomous theorem-discovery system. Rather, \LeanQIT{} supplies the typed objects, explicit side conditions, searchable dependencies, and public theorem endpoints needed for humans and proof agents to inspect, compose, and extend formal QIT. In this sense, \LeanQIT{} provides a digital foundation for auditable, scalable, and machine-checkable reasoning in quantum information theory.

\section*{Acknowledgment} This work was partially supported by the National Natural Science Foundation of China (Grant Nos.~92576114, 12447107) and the Guangdong Provincial Quantum Science Strategic Initiative (Grant Nos.~GDZX2403008, GDZX2503001, and GDZX2403001).
% ===================================================================
\clearpage
\appendix
\section{Selected Public Endpoint Wrappers}
\label{app:endpoint-wrappers}

The main text prioritizes declarations that expose formalization design or
proof composition.  This appendix records thin public wrappers that remain
useful as stable API endpoints but primarily forward already named results.
They introduce no additional mathematical claims.

\subsection{Fully quantum AEP endpoint}

The public AEP declaration obtains the nonempty subsystem instances carried by
a normalized bipartite state and then forwards the continuity and smooth-duality
assembly theorem discussed in Section~\ref{subsec:smooth-asymptotic}.

\begin{qitleancode}[title={Fully quantum AEP endpoint}]
theorem fullyQuantumAsymptoticEquipartitionProperty
    (\rho : State (Prod a b)) :
    QIT.asymptoticAEP_statement \rho := by
  letI : Nonempty a := by
    rcases \rho.nonempty with \langle x\rangle
    exact \langle x.1\rangle
  letI : Nonempty b := by
    rcases \rho.nonempty with \langle x\rangle
    exact \langle x.2\rangle
  exact \rho.asymptoticAEP_statement_of_traceEta_continuity_and_duality
    (State.SmoothMinRateUpperFromContinuity.afw \rho)
    (State.SmoothMaxRateFromMinDuality.smoothDuality \rho)
\end{qitleancode}
	
	\subsection{HSW source-facing package}
	
	The source-facing HSW endpoint pairs the operational capacity equality proved
	by the direct/converse squeeze with the epsilon-limit form of the regularized
	Holevo sequence.
	
\begin{qitleancode}[title={HSW public endpoint}]
namespace Channel

theorem hswClassicalCapacityTheorem_proved
    [Nonempty a] [Nonempty b] (N : Channel a b) :
    N.classicalCapacity = N.regularizedHolevoInformation \wedge
      (\forall \epsilon : \mathbbR, 0 < \epsilon \to
        \exists N0 : \mathbbN, \forall n : \mathbbN, n \ge N0 \to
          |N.blockHolevoInformation n / (n : \mathbbR) -
            N.regularizedHolevoInformation| < \epsilon) := by
  exact \langle classicalCapacity_eq_regularizedHolevoInformation N,
    regularizedHolevoInformation_limit N\rangle

end Channel
\end{qitleancode}
	
	\Needspace{28\baselineskip}
	\subsection{Entanglement-assisted asymptotic endpoints}
	
	The first wrapper combines the Petz direct input with the sandwiched-R\'enyi
	asymptotic upper input to expose ordinary and strong-converse upper bounds.
	
\begin{qitleancode}[title={EA asymptotic upper endpoint}]
namespace Channel

theorem entanglementAssisted_asymptoticUpperBounds_of_sandwichedLimit
    [Nonempty a] [Nonempty b] :
    N.IsEntanglementAssistedClassicalRateUpperBound
        N.entanglementAssistedInformation \wedge
      N.IsStrongConverseEntanglementAssistedClassicalRate
        N.entanglementAssistedInformation \wedge
      N.entanglementAssistedClassicalCapacity \le
        N.entanglementAssistedInformation \wedge
      N.strongConverseEntanglementAssistedClassicalCapacity \le
        N.entanglementAssistedInformation := by
  exact N.entanglementAssisted_asymptoticUpperBounds_of_sourceAsymptoticUpperInput
    (N.entanglementAssistedInformation_isAchievable_of_oneShotPetzLowerBound)
    (N.entanglementAssisted_sourceAsymptoticUpperInput_of_sandwichedLimit)

end Channel
\end{qitleancode}
	
	The final wrapper instantiates the abstract capacity squeezes with the same
	direct input and the source-consistent converse-witness family.
	
\begin{qitleancode}[title={EA capacity and strong-converse endpoint}]
namespace Channel

theorem entanglementAssisted_capacity_and_strongConverseCapacity_eq_information_of_sandwichedLimit
    [Nonempty a] [Nonempty b] :
    N.entanglementAssistedClassicalCapacity = N.entanglementAssistedInformation \wedge
      N.strongConverseEntanglementAssistedClassicalCapacity =
        N.entanglementAssistedInformation := by
  exact
    N.entanglementAssisted_capacity_and_strongConverseCapacity_eq_information_of_sourceConverseWitness
      (N.entanglementAssistedInformation_isAchievable_of_oneShotPetzLowerBound)
      (N.entanglementAssisted_sourceConverseWitnessFamily_of_sourceAsymptoticUpperInput
        (N.entanglementAssisted_sourceAsymptoticUpperInput_of_sandwichedLimit))

end Channel
\end{qitleancode}
	
	% ===================================================================
	\clearpage
	\bibliographystyle{unsrt}
	\bibliography{ref}
	
\end{document}

%% file: macro.tex
\usepackage{authblk} % package for author list
\usepackage[titletoc,title]{appendix}
\usepackage{hyperref}
\hypersetup{colorlinks=true,citecolor=blue,linkcolor=blue,filecolor=blue,urlcolor=blue,breaklinks=true}
\usepackage[dvipsnames]{xcolor}
\usepackage{color}
\usepackage[marginal]{footmisc}
\usepackage{url}
\usepackage{cite}
\usepackage{mathtools}
\usepackage{amsmath}
\usepackage{amssymb}
\usepackage[shortlabels]{enumitem}
\usepackage{graphicx,epic,eepic,epsfig,latexsym,verbatim}
\usepackage{dsfont}

\usepackage{framed}
\definecolor{shadecolor}{rgb}{0.9,0.9,0.9}
\usepackage{subcaption}
\usepackage{caption}
\usepackage{float}
\usepackage{tikz}
\usetikzlibrary{positioning}
\usetikzlibrary{plotmarks}
\usetikzlibrary{arrows.meta}
\usepackage{tcolorbox}
\usepackage{relsize}

\usepackage{pgfplots}
\pgfplotsset{compat=newest}
\usepgfplotslibrary{patchplots}

\definecolor{LightGray}{RGB}{220,220,220}
\definecolor{myred}{RGB}{236, 17, 0}
\definecolor{myblue}{RGB}{10, 88, 153}
\definecolor{myorange}{RGB}{236, 137, 0}
\definecolor{mygreen}{RGB}{26, 152, 81}

\usepackage{amsthm}
\newtheorem{theorem}{Theorem}[section]

\newtheorem{definition}[theorem]{Definition}

\def\squareforqed{\hbox{\rlap{$\sqcap$}$\sqcup$}}
\def\qed{\ifmmode\squareforqed\else{\unskip\nobreak\hfil
      \penalty50\hskip1em\null\nobreak\hfil\squareforqed
      \parfillskip=0pt\finalhyphendemerits=0\endgraf}\fi}
\def\endenv{\ifmmode\;\else{\unskip\nobreak\hfil
      \penalty50\hskip1em\null\nobreak\hfil\;
      \parfillskip=0pt\finalhyphendemerits=0\endgraf}\fi}

\newcounter{remark}

\newcounter{example}

\mathchardef\ordinarycolon\mathcode`\:
\mathcode`\:=\string"8000
\def\vcentcolon{\mathrel{\mathop\ordinarycolon}}
\begingroup \catcode`\:=\active
\lowercase{\endgroup
  \let :\vcentcolon
}

\usepackage{colortbl}
\usepackage{pifont}% http://ctan.org/pkg/pifont
\definecolor{Gray}{gray}{0.92}
\definecolor{Gray2}{gray}{0.75}
\definecolor{maroon}{cmyk}{0,0.87,0.68,0.32}
\usepackage{booktabs}
\usepackage{makecell}%To keep spacing of text in tables
\usepackage{diagbox}
\usepackage{multirow}

\newcommand{\tr}{\operatorname{Tr}}

\newcommand{\cN}{{\cal N}}

\newcommand{\RR}{{{\mathbb R}}}

\newcommand{\id}{{\operatorname{id}}}

\def\ve{\varepsilon}

\makeatletter
\newcommand*\rel@kern[1]{\kern#1\dimexpr\macc@kerna}
\newcommand*\widebar[1]{%
  \begingroup
  \def\mathaccent##1##2{%
    \rel@kern{0.8}%
    \overline{\rel@kern{-0.8}\macc@nucleus\rel@kern{0.2}}%
    \rel@kern{-0.2}%
  }%
  \macc@depth\@ne
  \let\math@bgroup\@empty \let\math@egroup\macc@set@skewchar
  \mathsurround\z@ \frozen@everymath{\mathgroup\macc@group\relax}%
  \macc@set@skewchar\relax
  \let\mathaccentV\macc@nested@a
  \macc@nested@a\relax111{#1}%
  \endgroup
}
\makeatother

\usepackage{algorithm}
\usepackage{algorithmic}

\usepackage{mathrsfs}
\newcommand{\Petz}{{\scriptscriptstyle  \rm P}}

\usepackage{stmaryrd}

\usepackage{pgfplots}